%% file: main.tex
\documentclass[a4paper,11pt]{article}
\usepackage{jheppub} 
\usepackage{tcolorbox}
\usepackage{xcolor}
\usepackage{lipsum}
\usepackage{tikz}
\usepackage[compat=1.0.0]{tikz-feynman}
\usepackage{graphicx}
\usepackage{subfig}
\usepackage{color}
\usepackage{appendix}
\usepackage{mathtools, nccmath, amssymb}
\allowdisplaybreaks[1]
\newcounter{desccount}
\newcommand{\descitem}[1]{%
  \item[#1] \refstepcounter{desccount}\label{#1}
}
\newcommand{\descref}[1]{\hyperref[#1]{#1}}
\title{\boldmath One-particle irreducibility of Vilkovisky-DeWitt effective action}

\author[a]{Sukanta Panda,}
\author[a]{Abbas Tinwala,}
\author[a]{Ketankumar Jadav}

\affiliation[a]{Department of Physics, Indian Institute of Science Education and Research Bhopal - 462066, India}

\emailAdd{abbas18@iiserb.ac.in}
\emailAdd{sukanta@iiserb.ac.in}
\emailAdd{jadav18@iiserb.ac.in}

\abstract{The effective action formalism introduced by Vilkovisky and later modified by DeWitt is completely covariant suitable for obtaining an effective action that is independent of the parametrization of the quantum fields. Among a few basic properties of an effective action, one property is that it can be written as a sum of one-particle irreducible diagrams. In this work, we verify this property up to three-loop order for Vilkovisky-Dewitt effective action (VDEA) by solving the equations satisfied by VDEA written down in the original parametrization for non-gauge theories.}

\begin{document}
\maketitle
\flushbottom

\section{Vilkovisky-DeWitt effective action in a nutshell}\label{Intro}
   \input{Introduction}

\section{Methods of computing VDEA}\label{SecII}   
    \input{Methods}

\section{Perturbative expansion of the VDEA}\label{SecIII}
\input{Section3}

\section{One-particle irreducibility of VDEA at two-loop}\label{SecIV}

\input{Section4}

\section{One-particle irreducibility of VDEA at three-loop}\label{SecV}
\input{Section5}

\section{Conclusion}\label{conclusion}
\input{Conclusion}

\acknowledgments
The calculations in this paper have been carried out in MATHEMATICA using the xAct packages xTensor \cite{xTensor}, and xPert \cite{xPert}. This work is partially supported by the DST (Govt. of India) Grant No. SERB/PHY/2021057.

\appendix
\input{Appendix}

\bibliographystyle{JHEP}
 \bibliography{biblio.bib}

\end{document}

%% file: Introduction.tex
We give only a brief introduction to the Vilkovisky-DeWitt effective action (VDEA) in this section. Emphasis has particularly been laid on the methods to compute VDEA following which we focus on the question of its one-particle irreducibility. We suggest to interested readers the detailed and pedagogical review found in \cite{tomsbook} and \cite{Odintsovbook} (Chapter 6 of \cite{tomsbook} contains most of the details about quantities that bear only a brief mention in this paper).

Vilkovisky has shown that the standard effective action is dependent on the parametrization chosen for the quantum fields \cite{Vilkovisky:1984st}. It is not difficult to see this for the standard one-loop effective action given by

\begin{align}
    \Gamma^{(1)}_\text{standard} = \frac{i\hbar}{2}\text{Tr Log }\frac{\delta^2 S[\bar\varphi]}{\delta\bar\varphi^i\bar\varphi^j},
\end{align}

where $\bar\varphi^i = \bar\varphi^I(x)$ are a set of mean fields. Here and throughout the paper we follow DeWitt's condensed notation described briefly in the Appendix \ref{calcCinv}.  It is obvious that the quantity $\frac{\delta^2 S[\bar\varphi]}{\delta\bar\varphi^i\bar\varphi^j}$ does not transform covariantly (as a tensor) if the fields are redefined: $\bar\varphi^i\rightarrow\bar\varphi'^i(\bar\varphi^j)$. Thus, even though the classical action $S[\varphi]$ is a scalar functional of the fields $\varphi^i$, the effective action is not a scalar functional of the mean fields. 

Vilkovisky proposed a formalism that would yield a covariant effective action by construction. In his proposal, he considered the field variables $\varphi^i$ as coordinates of a point in a Riemannian manifold $\mathcal{M}$ endowed with a metric tensor $g_{ij}[\varphi^k]$, with affine connections $\Gamma^i_{jk}$ to define differentiation of the quantities which are functional of the field variables. Furthermore, for a geodesic connecting the points with coordinates $\varphi^i$ and $\varphi'^i$, a vector quantity, $\sigma^i[\varphi,\varphi']$ was introduced that is tangent to the geodesic at the point with coordinate $\varphi$. This quantity naturally replaced the coordinate difference $\varphi^i-\varphi'^i$, which does not transform like a vector in the standard formalism. With the introduction of a covariant formalism, the equation satisfied by Vilkovisky's effective action is

\begin{align}\label{vdea1}
    e^{\frac{i}{\hbar}\tilde\Gamma[\bar\varphi,\varphi_*]} = \int \mathcal{D}\varphi \ \exp{\frac{i}{\hbar}\left\{S[\varphi]+\frac{\delta\hat\Gamma[\sigma^i[\varphi_*,\bar\varphi],\varphi_*]]}{\delta\sigma^i[\varphi_*,\bar\varphi]}\left(\sigma^i[\varphi_*,\bar\varphi]-\sigma^i[\varphi_*,\varphi]\right)\right\}}.
\end{align}

The quantities found in this equation can be explained as follows.
\begin{itemize}
    \item The point $\varphi_*^i$ is to be considered as the origin for the normal coordinates system formed locally on the manifold.
    \item The mean field $\bar\varphi$ is defined through the equation $\sigma^i[\varphi_*,\bar\varphi] = \langle \sigma^i[\varphi_*,\varphi]\rangle$ where the angular brackets denote the average defined in the path integral formalism
    \begin{align}
        \langle F[\varphi]\rangle = e^{-\frac{i}{\hbar}\tilde\Gamma[\bar\varphi,\varphi_*]}\int \mathcal{D}\varphi \ F[\varphi] \exp{\frac{i}{\hbar}\left\{S[\varphi]+\frac{\delta\hat\Gamma[\sigma^i[\varphi_*,\bar\varphi],\varphi_*]]}{\delta\sigma^i[\varphi_*,\bar\varphi]}\left(\sigma^i[\varphi_*,\bar\varphi]-\sigma^i[\varphi_*,\varphi]\right)\right\}}
    \end{align}

    \item The Vilkovisky effective action $\Gamma$ may be regarded either as a function of the original mean fields $\bar\varphi^i$ and $\varphi_*^i$, or as a function of $\sigma^i[\varphi_*,\bar\varphi]$ and $\varphi_*$, i.e. $\Gamma \equiv \tilde\Gamma[\bar\varphi,\varphi_*]\equiv\hat\Gamma[\sigma^i[\varphi_*,\bar\varphi],\varphi_*]$. Note that we have placed a tilde over $\Gamma$ if it is regarded as a function of the original mean fields $\bar\varphi^i$ and we have placed a hat over $\Gamma$ if it is regarded as a function of $\sigma^i[\varphi_*,\bar\varphi]$ (This notation can be found in \cite{Rebhan:1986wp, Rebhan:1987cd}).
\end{itemize}

The equation (\ref{vdea1}) is covariant because $\sigma^i[\varphi_*,\varphi]$ transforms like a vector at $\varphi^i_*$ and as a scalar at $\varphi^i$. The covariance of (\ref{vdea1}) guarantees that the effective action obtained from it is independent of the parametrization of the fields. 

The paper is organized as follows. In the next section (\ref{SecII}), we discuss ways to compute VDEA for non-gauge theories after which we turn towards the question of its one-particle irreducibility. In Section (\ref{SecIII}), we present the solution of (\ref{vdea1}) in the form of a series in powers of $\hbar$. In Section (\ref{SecIV}) and (\ref{SecV}) we show that VDEA is one-particle irreducible at two-loop and three-loop order respectively.

%% file: Methods.tex
We focus on the ways to solve \ref{vdea1} in this section. Let us define $v^i = \sigma^i[\varphi_*,\bar\varphi]$ for brevity. We also drop the argument of $\sigma^i[\varphi_*,\varphi]$ assuming it is always a function of $\varphi^i$ and $\varphi_*^i$. If $\sigma^i$ carries a different argument elsewhere we will restore them to avoid any confusion. The variational derivative of $\hat\Gamma[v^i,\varphi_*]$ in (\ref{vdea1}) can be expressed in terms of $\tilde\Gamma[\bar\varphi,\varphi_*]$ using simple chain rule to obtain

\begin{align}\label{vdea2}
    e^{\frac{i}{\hbar}\tilde\Gamma[\bar\varphi,\varphi_*]} = \int \mathcal{D}\varphi \ \exp{\frac{i}{\hbar}\left\{S[\varphi]+\frac{\tilde\Gamma[\bar\varphi,\varphi_*]}{\delta\bar\varphi^j}v^j{}_{,i}\left(v^i-\sigma^i\right)\right\}}.
\end{align}

It was shown in \cite{Burgess:1987zi} that although $\Gamma[\bar\varphi,\varphi_*]$ in (\ref{vdea1}) depends on $\varphi_*$, any choice of $\varphi_*$ defines an equally good effective action. DeWitt \cite{Batalin:1987fw} has further shown that the choice $\varphi^i_* = \bar\varphi^i$ is easiest for the calculation of $\tilde\Gamma[\bar\varphi, \varphi_*]$. The ``Vilkovisky-DeWitt" effective action or ``VDEA" is then defined by setting $\varphi_*^i$ equal to $\bar\varphi^i$ in the Vilkovisky effective action

\begin{align}
    \tilde\Gamma[\bar\varphi,\varphi_*]|_{\varphi_*=\bar\varphi} = \Gamma[\bar\varphi].
\end{align}

It turns out there are at least two methods to compute VDEA from either (\ref{vdea1}) or (\ref{vdea2}): 

\begin{description}
    \item[Method 1:] To solve (\ref{vdea1}) without invoking the identification $\varphi_*^i=\bar\varphi^i$ and carrying out the perturbative expansion of both sides of (\ref{vdea1})  in terms of the variable $\sigma^i$ about $\sigma^i=v^i$. The limit $\varphi_*^i = \bar\varphi^i$ is taken after computing $\hat\Gamma[v^i;\varphi_*]$

    \item[Method 2:] To take the limit $\varphi_*^i=\varphi^i$ in (\ref{vdea2}) itself and then proceed with the perturbative expansion of both sides of (\ref{vdea2}) in terms of the variable $\sigma^i$ about $\sigma^i=0$ using the fact that $v^i=0$ when $\varphi_*^i=\varphi^i$. The variable $\sigma^i$ is then expanded in normal coordinates followed by the path integral carried out in the original parametrization $\varphi^i$.
\end{description}

Let us discuss these methods briefly. The first method results in a straightforward use of (\ref{vdea1}) with the requirement of changing the path integral variable: $\varphi^i\rightarrow\sigma^i$. This introduces a Jacobian which must be taken into account while performing the perturbative expansion. For perturbative expansion, the covariant Taylor expansion of the classical action must be carried out on the R.H.S. of (\ref{vdea1}) which involves variational derivatives of the classical action $S[\bar\varphi]$ wrt. $v^i[\varphi_*,\bar\varphi]$ because we expand around $\sigma^i=v^i$

\begin{align}\label{covtaylor}
  S[\varphi]=S[\bar\varphi] + \sum_{n=1}^\infty\frac{1}{n!}\frac{\delta^n \hat S[v^i,\varphi_*]}{\delta v^{i_1}\delta v^{i_2} ... \delta v^{i_n}}(\sigma^{i_1}-v^{i_1})(\sigma^{i_2}-v^{i_2})...(\sigma^{i_n}-v^{i_n}).
\end{align}

where we considered $S[\bar\varphi] = \hat S[v^i,\varphi_*]$. These functional derivatives appear once we have computed $\tilde\Gamma[\bar\varphi;\varphi_*]$ up to a desired order and are in general given by an infinite sum of the conventional functional derivatives that makes them impossible to compute.

\begin{align}\label{vder}
    \frac{\delta^n \hat S[v^i;\varphi_*]}{\delta v^{i_1}\delta v^{i_2} ... \delta v^{i_n}} = (-1)^nS_{;(i_1i_2...i_n)}[\bar\varphi] + \sum_{m=n+1}^\infty\frac{(-1)^m}{(m-n)!}S_{;(i_1i_2...i_m)}v^{i_{n+1}}v^{i_{n+2}}...v^{i_m},
\end{align}

    where a semicolon denotes covariant functional derivatives that are to be computed using the field space connections. However when we carry out the final step of the calculation, i.e. take the limit $\varphi_*^i\rightarrow\bar\varphi^i$, we observe that the infinite sum part of (\ref{vder}) vanishes leaving behind symmetrized covariant functional derivatives of the classical action wrt. the mean fields in the original parametrization. These can be computed as usual for any theory using the field space connections.

The second method can be used after we rewrite $(\ref{vdea2})$ completely in terms of $\varphi$ and $\bar\varphi$ by taking the limit $\varphi_*=\varphi$. This requires us to compute the following limit in (\ref{vdea2})

\begin{align}
    \frac{\delta\tilde\Gamma[\bar\varphi,\varphi_*]}{\delta \bar\varphi^j}\Big|_{\varphi_*=\bar\varphi}.
\end{align}
This is not simply $\frac{\Gamma[\bar\varphi]}{\delta \bar\varphi^j}$ because the derivative must be computed before taking the limit inside $\Gamma[\bar\varphi,\varphi_*]$. To calculate the limit properly we require the following identity which can be derived from (\ref{vdea2})

\begin{align}\label{3id1}
    \frac{\delta\tilde\Gamma[\bar\varphi;\varphi_*]}{\delta \varphi^i_*} = \frac{\delta\tilde\Gamma[\bar\varphi;\varphi_*]}{\delta \bar\varphi^j}(v^j{}_{;i}-C^j{}_{i}[\bar\varphi]),
\end{align}
where 
\begin{align}
    C^i{}_{j}[\bar\varphi]=\langle\sigma^j{}_{;i}[\bar\varphi,\varphi]\rangle.
\end{align}
After taking the limit $\varphi_*=\bar\varphi$ in (\ref{3id1}) we obtain the following identity using (\ref{vij})

\begin{align}\label{3id2}
    \frac{\delta\tilde\Gamma[\bar\varphi,\varphi_*]}{\delta \bar\varphi^i}\Big|_{\varphi_*=\bar\varphi} = C^{-1i}{}_{j}[\bar\varphi]\times\lim_{\varphi_*\rightarrow\bar\varphi}\left(\frac{\delta\tilde\Gamma[\bar\varphi;\varphi_*]}{\delta\varphi_*^j}+\frac{\delta\tilde\Gamma[\bar\varphi;\varphi_*]}{\delta\bar\varphi^j}\right).
\end{align}

Finally, we note that
\begin{align}
    \delta\Gamma[\bar\varphi]
    &=\lim_{\varphi_*=\bar\varphi}\delta\tilde\Gamma[\varphi_*;\varphi]\nonumber\\
    &=\lim_{\varphi_*=\bar\varphi}\left(\frac{\delta\tilde\Gamma[\bar\varphi;\varphi_*]}{\delta\varphi_*^i}\delta\varphi_*^i+\frac{\delta\tilde\Gamma[\bar\varphi;\varphi_*]}{\delta\bar\varphi^i}\delta\bar\varphi^i\right)\nonumber\\
    &=\lim_{\varphi_*=\bar\varphi}\left(\frac{\delta\tilde\Gamma[\bar\varphi;\varphi_*]}{\delta\varphi_*^i}+\frac{\delta\tilde\Gamma[\bar\varphi;\varphi_*]}{\delta\bar\varphi^i}\right)\delta\bar\varphi^i\label{3id3}
\end{align}
Using (\ref{3id3}) in (\ref{3id2}) we find

\begin{align}
    \frac{\delta\tilde\Gamma[\bar\varphi,\varphi_*]}{\delta \bar\varphi^i}\Big|_{\varphi_*=\bar\varphi} = C^{-1i}{}_{j}[\bar\varphi]\frac{\delta\Gamma[\bar\varphi]}{\delta\bar\varphi^j}.
\end{align}
Now we can rewrite (\ref{vdea2}) as

\begin{align}\label{vdea2re}
    e^{\frac{i}{\hbar}\Gamma[\bar\varphi]} = \int \mathcal{D}\varphi \ \exp{\frac{i}{\hbar}\left\{S[\varphi]-\sigma^i[\bar\varphi,\varphi]C^{-1j}{}_{,i}[\bar\varphi]\Gamma_{,j}[\bar\varphi]\right\}}
\end{align}

This equation for $\Gamma[\bar\varphi]$ is supplemented by an equation for $C^i{}_j[\bar\varphi]$ that reads

\begin{align}\label{Ceq}
    C^i{}_j[\bar\varphi]=e^{-\frac{i}{\hbar}\Gamma[\bar\varphi]}\int \mathcal{D}\varphi \ \sigma^i{}_{;j}[\bar\varphi,\varphi]\exp{\frac{i}{\hbar}\left\{S[\varphi]-\sigma^i[\bar\varphi,\varphi]C^{-1j}{}_{,i}[\bar\varphi]\Gamma_{,j}[\bar\varphi]\right\}}.
\end{align}
(\ref{vdea2re}) and (\ref{Ceq}) thus form a set of coupled integro-differential equations for $\Gamma[\bar\varphi]$ and $C^i{}_j[\bar\varphi]$. The computation of the VDEA then involves solving both (\ref{vdea2re}) and (\ref{Ceq}) simultaneously up to any desired order. Since the limit $\varphi_* = \bar\varphi$ has already been taken in (\ref{vdea2re}) and (\ref{Ceq}) we just need to use the covariant Taylor series expansion about $\sigma^i=0$ with all the vertices being the conventional ones, i.e. $S_{;(i_1i_2...i_n)}$. 

Let us discuss the merits and demerits of using either of these methods. We observe that method (1) is relatively simple since we need to solve only one equation whereas method (2) involves coupled integro-differential equations which are much more complicated and tedious to solve as we will see in the following sections. The only problem with method (1) is that the change in the variable of the path integral introduces a jacobian which as shown in \cite{Rebhan:1987cd} starts to contribute from two-loop order onwards. As usual, the jacobian can be dealt with easily by introducing the usual Fadeev-Popov ghosts in the path integral \cite{Rebhan:1987cd}. The only merit of method (2) is that it allows us to carry out the perturbative expansion in the original parametrization starting from (\ref{vdea2re}) which involves only conventional vertices from the beginning. This however, does not seem at all to be an advantage over method (1) since although method (1) involves at an intermediate step an infinite sum of vertices finally it is the conventional vertices that are to be computed which can be done easily for any interacting theory. The result of this discussion is that method (1) is more of a practical use for computing VDEA. It is simple to show that VDEA is a sum of one-particle irreducible diagrams through the use of method (1) \cite{Rebhan:1987cd}. For consistency, it becomes imperative to show that this holds even if one chooses to use method (2) which is what forms the main motive of this paper. We also note that up to one-loop order there is no difference between the two methods discussed since $C^{-1i}{}_{j}[\bar\varphi]$ only makes non-trivial contributions when VDEA is computed beyond one-loop (See \ref{1l}). In the next section, we perturbatively expand VDEA in integral powers of $\hbar$.

%% file: Section3.tex
The procedure to solve (\ref{vdea2}) differs from the standard one only in that a covariant Taylor expansion of $S$ is employed rather than the usual Taylor expansion. This covariant Taylor expansion of $S$ is performed about $\sigma^i=v^i$ in the RHS of (\ref{vdea2}) using (\ref{covtaylor}). We may also consider the effective action and $C^{-1i}{}_{j}[\bar\varphi]$ as an infinite series expressed in powers of $\hbar$

\begin{align}
    \Gamma[\bar\varphi]=\sum_{n=0}^\infty\hbar^n\Gamma^{(n)}[\bar\varphi]
\end{align},

\begin{align}\label{Cinvexp}
    C^{-1i}{}_{j}[\bar\varphi] = \sum_{n=0}^\infty\hbar^n C_n^{-1i}{}_{j}[\bar\varphi]
\end{align}.

After the covariant Taylor expansion is performed about $\sigma^i=v^i$ on the RHS of (\ref{vdea2}) we immediately take the limit $\varphi_* = \bar\varphi$ to obtain 

\begin{align}\label{rhs1}
    \int \mathcal{D}\varphi \  \exp{\frac{i}{\hbar}\Big\{S[\bar\varphi] + \sum_{n=0}^\infty \frac{\delta^nS[\bar\varphi]}{\delta\sigma^{i_1}\delta\sigma^{i_2}...\delta\sigma^{i_n}}\sigma^{i_1}\sigma^{i_2}...\sigma^{i_n}-\sigma^i\sum_{k=0}^\infty\hbar^kC_k^{-1i}{}_{j}[\bar\varphi]\sum_{m=0}^\infty\hbar^m\Gamma^{(m)}[\bar\varphi]\Big\}},
\end{align}

where we used the fact that $v^i = 0$ when $\varphi^i_* = \bar\varphi^i$. If a coordinate system is constructed locally about the origin $\bar\varphi^i$, then the quantity $\sigma^i[\bar\varphi,\varphi]$ can be expanded in normal coordinates $\eta^i$ as follows

\begin{align}\label{sigmaexp}
    \sigma^i[\bar\varphi,\varphi]=-\eta^i + \sum_{n=1}^\infty\sigma^i_{(n)},
\end{align}
where the first three terms in the expansion are 
\begin{align}
    &\eta^i = \varphi^i-\bar\varphi^i,\\
    &\sigma^i_{(2)} = -1\frac{1}{2}\Gamma^i_{jk}\eta^j\eta^k,\\
    &\sigma^i_{(3)} = -\frac{1}{6}(\Gamma^i_{jk,l}+\Gamma^i_{jm}\Gamma^m_{kl})\eta^j\eta^k\eta^l.
\end{align}

Using the normal coordinate expansion of $\sigma^i[\bar\varphi,\varphi]$ and redefining the path integral variable $\varphi^i\rightarrow\hbar\eta^i$ in (\ref{rhs1}) we obtain

\begin{align}
    \int \mathcal{D}\eta \exp{\frac{i}{\hbar}\Big\{S[\bar\varphi]+\hbar A_2[\bar\varphi]+\hbar\sqrt{\hbar}A_3[\bar\varphi]+\hbar^2A_4[\bar\varphi]+\hbar^2\sqrt{\hbar}A_5[\bar\varphi]+\hbar^3A_6[\bar\varphi]+\mathcal{O}(\hbar^{7/2})\Big\}},
\end{align}

where the series has been terminated at cubic power of $\hbar$ to enable us to examine VDEA up to three loops. The various functions appearing in the expression above read

\begin{align}
    &A_2[\bar\varphi] = \frac{1}{2}S_{;ij}\eta^i\eta^j,\\
    &A_3[\bar\varphi] = \frac{1}{6}S_{;(ijk)}\eta^i\eta^j\eta^k-S_{,i}C_1^{-1i}{}_{j}[\bar\varphi]\eta^j-\Gamma^{(1)}_{,i}[\bar\varphi]\eta^i-S_{;ij}\eta^i\sigma^j_{(2)},\\
    &A_4[\bar\varphi] = \frac{1}{24}S_{;(ijkl)}\eta^i\eta^j\eta^k\eta^l+\Gamma^{(1)}_i\sigma^i_{(2)}+S_{,i}C_1^{-1i}{}_{j}[\bar\varphi]\sigma_{(2)}^j-\frac{1}{2}S_{;(ijk)}\eta^i\eta^j\sigma^k_{(2)}\nonumber\\
    &\hspace{12mm}+\frac{1}{2}S_{;ij}\sigma^i_{(2)}\sigma^j_{(2)}-S_{;ij}\eta^i\sigma_{(3)}^j,\\
    &A_5[\bar\varphi] = \frac{1}{120}S_{;(ijklm)}\eta^i\eta^j\eta^k\eta^l\eta^m-\Gamma^{(2)}_{,i}\eta^i-\Gamma^{(1)}_{,i}C_1^{-1i}{}_{j}[\bar\varphi]\eta^j-S_{,i}C_2^{-1i}{}_{j}[\bar\varphi]\eta^j\nonumber\\
    &\hspace{12mm}+\frac{1}{2}S_{;(ijk)}\eta^i\sigma_{(2)}^j\sigma_{(2)}^k-\frac{1}{6}S_{;(ijkl)}\eta^i\eta^j\eta^k\sigma_{(2)}^l+\Gamma^{(1)}_{,i}\sigma_{(3)}^i+S_{,i}C_1^{-1i}{}_{j}[\bar\varphi]\sigma_{(3)}^j\nonumber\\
    &\hspace{12mm}+S_{;ij}\sigma^i_{(2)}\sigma^j_{(3)}-\frac{1}{2}S_{;(ijk)}\eta^i\eta^j\sigma^k_{(3)}-S_{;ij}\eta^i\sigma^j_{(4)},\\
    &A_6[\bar\varphi] = \frac{1}{720}S_{;(ijklmn)}\eta^i\eta^j\eta^k\eta^l\eta^m\eta^n+\Gamma^{(2)}_i\sigma_{(2)}^i + \Gamma^{(1)}_{,i}C_1^{-1i}{}_{j}[\bar\varphi]\sigma_{(2)}^j+ S_{,i}C_2^{-1i}{}_{j}[\bar\varphi]\sigma_{(2)}^j\nonumber\\
    &\hspace{12mm}-\frac{1}{6}S_{;(ijk)}\sigma^i_{(2)}\sigma^j_{(2)}\sigma^k_{(2)}+\frac{1}{6}S_{;(ijkl)}\eta^i\eta^j\sigma^k_{(2)}\sigma^l_{(2)}+\frac{1}{12}S_{;(ijkl)}\eta^i\eta^j\sigma^k_{(2)}\sigma^l_{(2)}\nonumber\\
    &\hspace{12mm}-\frac{1}{24}S_{;(ijklm)}\eta^i\eta^j\eta^k\eta^l\sigma^m_{(2)}+\frac{1}{2}S_{;ij}\sigma^i_{(3)}\sigma^j_{(3)}+S_{;(ijk)}\eta^i\sigma^j_{(2)}\sigma^k_{(3)}\nonumber\\
    &\hspace{12mm}-\frac{1}{6}S_{;(ijkl)}\eta^i\eta^j\eta^k\sigma^l_{(3)}+\Gamma^{(1)}_{,i}\sigma^i_{(4)}+ S_{,i}C_1^{-1i}{}_{j}[\bar\varphi]\sigma_{(4)}^j+S_{;ij}\sigma^i_{(2)}\sigma^j_{(4)}\nonumber\\
    &\hspace{12mm}-\frac{1}{2}S_{;(ijk)}\eta^i\eta^j\sigma^k_{(4)}-S_{;ij}\eta^i\sigma^j_{(5)}.
\end{align}

Since the LHS and RHS of (\ref{vdea2}) are both polynomials of $\hbar$, the coefficient of each term in the polynomials can be compared on both sides of (\ref{vdea2}) to obtain VDEA in powers oh $\hbar$. Up to the third order in $\hbar$ (or three-loop order) we obtain
\begin{align}
    &\Gamma^{(0)} = S[\bar\varphi],\\
    &\Gamma^{(1)} = \frac{i\hbar}{2}\text{Tr Log}S_{;ij}\label{1l},\\
    &\Gamma^{(2)} = \hbar^2\Big\langle A_4 + \frac{i}{2} A_3^2\Big\rangle\label{2l},\\
    &\Gamma^{(3)} = \hbar^3\Big\langle A_6 + \frac{i}{2}A^2_4+iA_3A_5-\frac{1}{2}A^2_3A_4-\frac{i}{24}A_3^4\Big\rangle + \frac{i\hbar^3}{2}\Big\langle iA_4-\frac{1}{2}A_3^2\Big\rangle^2\label{3l},
    \end{align}

    where the angular brackets denote averages computed in the path-integral formalism, for example
    \begin{align}
        \langle F[\eta]\rangle =\frac{\int \mathcal{D}\eta F[\eta]\exp{\Big\{\frac{i}{2}S_{;ij}\eta^i\eta^j\Big\}}}{\int \mathcal{D}\eta\exp{\Big\{\frac{i}{2}S_{;ij}\eta^i\eta^j\Big\}}}. 
    \end{align}

In the following sections, we examine the one-particle irreducibility of VDEA up to three-loop order while choosing method (2) for the computations. For brevity, we will use the words ``1PI" and ``1PR" in place of one-particle irreducible and one-particle reducible respectively in the following sections.

%% file: Section4.tex
    Let us examine the two-loop VDEA (\ref{2l})
    
    \begin{align}
        \Gamma^{(2)} = \hbar^2\Big\langle A_4 + \frac{i}{2} A_3^2\Big\rangle.
    \end{align}
    
    The term $\langle A_4\rangle$ cannot contain any 1PR diagrams since the entire functional is defined at only one space-time point. This is because even though each Latin index over which the terms are summed over carries a unique set of field indices and a space-time coordinate, the ordinary functional derivatives of $S$ and the connection terms, $\Gamma^i_{jk}$, involve Dirac delta functions, reducing the integral over only one space-time coordinate. The following example is sufficient to make this point clear. Given the identifications: $i \leftrightarrow (I,y)$, $j \leftrightarrow (J,z)$ and $k \leftrightarrow (K,w)$ we find

    \begin{align}\label{example1}
        S_{;ij}\eta^i\eta^j &= S_{,ij}\eta^i\eta^j - \Gamma^k_{ij}S_{,k}\eta^i\eta^j\nonumber\\
        &=\int dv_y dv_z \frac{\delta^2S[\bar\varphi(x)]}{\delta \bar\varphi^I(y)\delta\bar\varphi^J(z)}\eta^I(y)\eta^J(z) - \int dv_y dv_z dv_w \Gamma^K_{IJ}(w,y,z)\frac{\delta S[\bar\varphi(x)]}{\delta \bar\varphi^K(w)}\eta^I(y)\eta^J(z),
    \end{align}
where we have used DeWitt notation \ref{notation}. The ordinary functional derivatives of $S$ may be considered to have in general the following form,

\begin{align}\label{varSform}
    &\frac{\delta^2S[\bar\varphi(x)]}{\delta \bar\varphi^I(y)\delta\bar\varphi^J(z)} = \int dv_x (\hat L_2)_{IJ}(x)\{\tilde\delta(x,y)\tilde\delta(x,z)\},\\
    &\frac{\delta S[\bar\varphi(x)]}{\delta \bar\varphi^K(w)} = \int dv_x (\hat L_1)_K(x)\{\tilde\delta(x,y)\},
\end{align}
where $\hat (L_2)_{IJ}(x)$ and $\hat (L_1)_K(x)$ are some operators acting on the delta functions. The connection terms are also proportional to Dirac delta functions (see \cite{paper1, paper2, paper3}), which in general can be written as

\begin{align}\label{connform}
    \Gamma^K_{IJ}(w,y,z) = (\gamma(w,y,z)\hat O(w))^K_{IJ}\{\tilde\delta(w,y)\tilde\delta(w,z)\},  
\end{align}
where $(\gamma(w,y,z)\hat O(w))^K_{IJ}$ is some operator\footnote{The connection terms given in \cite{paper1, paper2, paper3} are simply proportional to Dirac delta functions. These connection terms were derived for gauge theories when Landau-DeWitt gauge is used and as such they assume a simple form making the calculation of VDEA easier. In general, however, the connection terms are proportional to derivative operators acting on the Dirac delta functions as shown in \cite{cho1989vilkovisky, cho1991vilkovisky, odintsov1993gaugeh}} acting on the delta functions within the curly brackets.

Using (\ref{varSform}) and (\ref{connform}) in (\ref{example1}), we find

    \begin{align}
        S_{;ij}\eta^i\eta^j = &\int dv_xdv_y dv_z (\hat L_2)_{IJ}(x)\{\tilde\delta(x,y)\tilde\delta(x,z)\}\eta^I(y)\eta^J(z) \nonumber\\
        &- \int dv_x dv_y dv_y dv_w  (\hat L_1)_K(x)\{\tilde\delta(x,w)\}(\gamma(w,y,z)\hat O(w))^K_{IJ}\{\tilde\delta(w,y)\tilde\delta(w,z)\}\eta^I(y)\eta^J(z)\nonumber\\
         =&  \int dv_x (\hat L_2)_{IJ}(x) \{\eta^I(x)\eta^J(x)\} - \int dv_x(\hat L_1)_K(x)\big[(\gamma(x)\hat O(x))^K_{IJ}\{\eta^I(x)\eta^J(x)\}\big].\label{example1a}
    \end{align}
The form of the operators and the way they act depends on the action and the field space metric and it is necessary to know these to evaluate (\ref{example1a}). However, it is sufficient for our analysis to know that if (\ref{varSform}) and (\ref{connform}) hold then the expression (\ref{example1a}) can be written as an integral over one space-time coordinate.

    The result of the preceding discussion is that 1PR diagrams can emerge only from $\langle A_3^2\rangle$ which can be written as an integral over two space-time coordinates. To write down $\langle A_3^2\rangle$ explicitly we require the functional derivative of one-loop effective action which can be evaluated using (\ref{1l}) as
    
    \begin{align}
    \Gamma^{(1)}_{,k} &= \frac{i\hbar}{2}S_{;ijk}G^{ij}\nonumber\\
    &=\frac{\hbar}{2}S_{;ijk}\langle\eta^i\eta^j\rangle\label{der1l}
\end{align}

where we used $iG^{ij} = \langle \eta^i\eta^j\rangle$ and the fact that the Green's function $G^{ij}$ satisfies, $S_{;ij}G^{jk} = \delta^k_i$. It is important to note that there is no symmetrization over indices in $S_{;ijk}$ in (\ref{der1l}) (a fact specifically pointed out in \cite{Rebhan:1986wp, Rebhan:1987cd}).

If the field space metric is assumed flat, then $\langle A_3^2\rangle$ reads,
\begin{align}\label{flat2l}
    \Big\langle\frac{1}{6}S_{,ijk}\eta^i\eta^j\eta^k - \frac{1}{2}S_{,ijk}\langle\eta^i\eta^j\rangle\eta^k\Big\rangle^2,
\end{align}

where the covariant functional derivatives become ordinary functional derivatives and symmetrization over indices becomes trivial since the ordinary functional derivatives commute. If we expand (\ref{flat2l}) and take the average we find

\begin{align}
    \langle A_3^2\rangle &= S_{,ijk}S_{,lmn}\Big(\frac{1}{36}(9\langle\eta^i\eta^j\rangle\langle\eta^k\eta^l\rangle\langle\eta^m\eta^n\rangle + 6\langle\eta^i\eta^l\rangle\langle\eta^j\eta^m\rangle\langle\eta^k\eta^n\rangle)\nonumber\\
    &-\frac{1}{2}\langle\eta^i\eta^j\rangle\langle\eta^k\eta^l\rangle\langle\eta^m\eta^n\rangle+\frac{1}{4}\langle\eta^i\eta^j\rangle\langle\eta^k\eta^l\rangle\langle\eta^m\eta^n\rangle\Big)\nonumber\\
    &=\frac{1}{6}S_{,ijk}S_{,lmn}\langle\eta^i\eta^l\rangle\langle\eta^j\eta^m\rangle\langle\eta^k\eta^n\rangle\label{flat2la}.
\end{align}

This expression can be represented in a diagrammatic form as

\begin{align}
    \langle A_3^2\rangle = \frac{1}{6}\begin{tikzpicture}[baseline={([yshift=-.5ex]current bounding box.center)}]
        \begin{feynman}
            \node [dot] (a) at (0,0);
            \node [dot] (b) at (1,0);
            \diagram*{(a)--[red, half right] (b) -- [red] (a) -- [red, half left] (b)};
        \end{feynman}
    \end{tikzpicture},
\end{align}
where the black dots are vertices $S_{,ijk}$ and the lines connecting any two vertices are the propagators  $\langle\eta^i\eta^j\rangle$. We see that the 1PR diagrams all cancel away when the field space metric is flat. 

For the case of a general field space metric the expression for $\langle A_3^2\rangle$ reads

\begin{align}\label{gen2l}
    \langle A_3^2\rangle = \Big\langle\frac{1}{6}S_{;(ijk)}\eta^i\eta^j\eta^k- \frac{1}{2}S_{;ijk}\langle\eta^i\eta^j\rangle\eta^k - S_{,i}C_1^{-1i}{}_{j}\eta^j-S_{;ij}\eta^i\sigma^j_{(2)}\Big\rangle^2.
\end{align}

A similar cancellation of 1PR diagrams does not happen here if we consider only the first two terms in the expression above since $S_{;(ijk)}$ is different from $S_{;ijk}$. But if we use the expression for $C_1^{-1i}{}_{j}$ (\ref{Cinv1}) we find

\begin{align}
    - \frac{1}{2}S_{;ijk}\langle\eta^i\eta^j\rangle\eta^k - S_{,i}C_1^{-1i}{}_{j}\eta^j &= - \frac{1}{2}S_{;ijk}\langle\eta^i\eta^j\rangle\eta^k - \frac{1}{3}S_{,l}R^l{}_{ikj}\langle\eta^i\eta^j\rangle\eta^k\nonumber\\
    &=-\frac{1}{2}S_{;(ijk)}\langle\eta^i\eta^j\rangle\eta^k.
\end{align}

The expression for $\langle A_3^2\rangle$ now becomes

\begin{align}
    \langle A_3^2\rangle = \Big\langle\Big(\frac{1}{6}S_{;(ijk)}\eta^i\eta^j\eta^k- \frac{1}{2}S_{;(ijk)}\langle\eta^i\eta^j\rangle\eta^k - S_{;ij}\eta^i\sigma^j_{(2)}\Big)^2\Big\rangle.
\end{align}

Let us define for brevity

\begin{align}\label{Ls}
    &L_1 = \frac{1}{6}S_{;(ijk)}\eta^i\eta^j\eta^k,\\
    &L_2 = - \frac{1}{2}S_{;(ijk)}\langle\eta^i\eta^j\rangle\eta^k,\\
    &L_3 =  - S_{;ij}\eta^i\sigma^j_{(2)}.
\end{align}

The symmetrization over indices in $L_2$ causes 1PR diagrams to cancel in $\langle(L_1+L_2)^2\rangle$ as it was shown for the case of a flat field-space metric. However due to $L_3$ there exists the following extra terms in $\langle A_3^2\rangle$.

\begin{align}\label{extra2}
    \Big\langle(S_{;ij}\eta^i\sigma^j_{(2)})^2-2\Big(\frac{1}{6}S_{;(ijk)}\eta^i\eta^j\eta^k-\frac{1}{2}S_{;(ijk)}\langle\eta^i\eta^j\rangle\eta^k\Big)(S_{;lm}\eta^l\sigma^m_{(2)})\Big\rangle.
\end{align}
Consider the first term in the expression above while focussing only on the 1PR diagrams.
\begin{align}\label{extra1}
    \frac{1}{4}S_{;ij}S_{;mn}\Gamma^j{}_{kl}\Gamma^n{}_{pq}\langle \eta^i\eta^k\eta^l\eta^m\eta^p\eta^q\rangle &= \frac{1}{4}S_{;ij}S_{;mn}\Gamma^j{}_{kl}\Gamma^n{}_{pq}\Big(\langle\eta^i\eta^k\rangle\langle\eta^l\eta^m\rangle\langle\eta^p\eta^q\rangle \nonumber\\&
    + \langle\eta^i\eta^k\rangle\langle\eta^l\eta^p\rangle\langle\eta^m\eta^q\rangle + \langle\eta^i\eta^k\rangle\langle\eta^l\eta^q\rangle\langle\eta^p\eta^m\rangle \nonumber\\&
    + \langle\eta^i\eta^l\rangle\langle\eta^k\eta^m\rangle\langle\eta^p\eta^q\rangle + \langle\eta^i\eta^l\rangle\langle\eta^k\eta^p\rangle\langle\eta^m\eta^q\rangle \nonumber\\&
    + \langle\eta^i\eta^l\rangle\langle\eta^k\eta^q\rangle\langle\eta^p\eta^m\rangle + \langle\eta^k\eta^l\rangle\langle\eta^i\eta^p\rangle\langle\eta^m\eta^q\rangle\nonumber\\&
    + \langle\eta^k\eta^l\rangle\langle\eta^i\eta^q\rangle\langle\eta^m\eta^p\rangle + \langle\eta^k\eta^l\rangle\langle\eta^i\eta^m\rangle\langle\eta^p\eta^q\rangle\Big).
\end{align}
There are six more terms in the expression above but those involve only 1PI diagrams. The first term in RHS of the above expression reads
\begin{align}
    \frac{1}{4}S_{;ij}S_{;mn}\Gamma^j{}_{kl}\Gamma^n{}_{pq}\langle\eta^i\eta^k\rangle\langle\eta^l\eta^m\rangle\langle\eta^p\eta^q\rangle.
\end{align}
If we use the equation satisfied by the Green's function, $S_{;ij}\langle\eta^i\eta^k\rangle = i\delta^k_j$ the expression above becomes

\begin{align}
    \frac{1}{4}S_{;mn}\Gamma^j{}_{jl}\Gamma^n{}_{pq}\langle\eta^l\eta^m\rangle\langle\eta^p\eta^q\rangle.
\end{align}

Since the connections are proportional to Dirac delta functions, those with similar indices would give, $\Gamma^j{}_{jl} \propto \delta^4(0)$. If we use the dimensional regularisation scheme then connection terms with repeated indices would be regularised to zero;

\begin{align}\label{res1}
    \frac{1}{4}S_{;mn}\Gamma^j{}_{jl}\Gamma^n{}_{pq}\langle\eta^l\eta^m\rangle\langle\eta^p\eta^q\rangle \rightarrow \delta^4(0) \rightarrow 0
\end{align}

This happens for the first eight terms in the RHS of (\ref{extra1}). The ninth term reads
\begin{align}\label{ninth}
    \frac{1}{4}S_{;jn}\Gamma^j{}_{kl}\Gamma^n{}_{pq}\langle\eta^k\eta^l\rangle\langle\eta^p\eta^q\rangle.
\end{align}
Following the example (\ref{example1a}) we see that (\ref{ninth}) can be written as an integral over one space-time coordinate and hence does not contain any 1PR diagram. The fact that the equation satisfied by Green's function can be used to turn an apparent 1PR diagram into a 1PI diagram can be illustrated clearly in the following diagrammatic way;
\begin{align}\label{res2}
    \begin{tikzpicture}[baseline={([yshift=-.5ex]current bounding box.center)}]
    \begin{feynman}
    \vertex [dot, label={[shift={(0.7,-0.1)}]$x$}] (a) at (-3,0) ;
    \node [dot, label={[shift={(1.0,0)}]\tiny$\langle\eta^i(x)\eta^j(y)\rangle$}] (b) at (-2,0);
    \node [dot] (c) at (0,0);
    \vertex [dot, label={[shift={(-0.7,-0.1)}]$y$}] (d) at (1,0);
    \diagram*{
    (a) -- [red, half right] (b) -- [red] (c) -- [red, half right] (d) -- [red, half right] (c) -- [red] (b) -- [red, half right] (a)
    };
    \end{feynman}
    \end{tikzpicture} \times S_{;jk}(y)\rightarrow \begin{tikzpicture}[baseline={([yshift=-.5ex]current bounding box.center)}]
    \begin{feynman}
    \vertex (a) at (0,0);
    \node [dot] (b) at (1,0);
    \vertex (c) at (2,0);
    \diagram*{(a) -- [red, half right] (b) -- [red, half right] (c) -- [red, half right] (b) -- [red, half right] (a)};
    \end{feynman}
    \end{tikzpicture}
\end{align}
We see that two categories of diagrams can be formed when vertices with $S_{;ij}$ are involved;
\begin{description}
    \descitem{Result 1}: When the loop carries the same index as the second-order covariant functional derivative of $S$ then a $\delta^4(0)$ term is obtained which is regularised to zero. (See (\ref{res1})).
   \descitem{Result 2}: When the propagator connecting two different vertices carries the same index as the second-order covariant functional derivative of $S$ then the propagator shortens to ``zero-length" while bringing together the two points separated by it. (See (\ref{res2})).
\end{description}

These results will be important for us to show one-particle irreducibility up to three-loop order in the next section.

Let us now examine the second and third terms in the RHS of (\ref{extra2}) focusing only on the 1PR diagrams. The second term reads
\begin{align}\label{second}
    \frac{1}{6}S_{;(ijk)}S_{;lm}\Gamma^m{}_{pq}\langle\eta^i\eta^j\eta^k\eta^l\eta^p\eta^q\rangle = \frac{1}{6}S_{;(ijk)}S_{;lm}\Gamma^m{}_{pq}\Big(3\langle\eta^i\eta^j\rangle\langle\eta^k\eta^l\rangle\langle\eta^p\eta^q\rangle + 6\langle\eta^i\eta^j\rangle\langle\eta^k\eta^p\rangle\langle\eta^l\eta^q\rangle\Big). 
\end{align}
 Six additional terms that involve only 1PI diagrams have been ignored in the expression above. The third term in the RHS of (\ref{extra2}) reads
 \begin{align}\label{third}
     -\frac{1}{2}S_{;(ijk)}S_{;lm}\Gamma^m{}_{pq}\langle\eta^i\eta^j\rangle\langle\eta^k\eta^l\eta^p\eta^q\rangle = -\frac{1}{2}S_{;(ijk)}S_{;lm}\Gamma^m{}_{pq}\langle\eta^i\eta^j\rangle\Big(\langle\eta^k\eta^l\rangle\langle\eta^p\eta^q\rangle + 2\langle\eta^k\eta^p\rangle\langle\eta^l\eta^q\rangle\Big).
 \end{align}
 On adding (\ref{second}) and (\ref{third}), we find that all 1PR diagrams cancel away.

 Thus, we have shown that $\langle A_3^2\rangle$ involves only 1PI diagrams, and hence VDEA contains only 1PI diagrams up to two-loop order.

%% file: Section5.tex
Let us examine VDEA at three-loop order (\ref{3l}). In this section, we will only focus on terms that would give rise to 1PR diagrams;

\begin{align}
    \Gamma^{(3)}_\text{1PR} = \hbar^3\Big\langle \frac{i}{2}A^2_4+iA_3A_5-\frac{1}{2}A^2_3A_4-\frac{i}{24}A_3^4\Big\rangle 
\end{align}
 Let us examine each term individually.
\subsection{$\langle A^4_3 \rangle$}\label{SecV1}
We expand $\langle A^4_3 \rangle$ in terms of $L_1$, $L_2$ and $L_3$ (See (\ref{Ls}));

\begin{align}\label{3l1}
    \langle A^4_3 \rangle &= \langle(L_1 + L_2 + L_3)^4\rangle\nonumber\\ &= \langle(L_1+L_2)^4\rangle + 4\langle(L_1+L_2)^3L_3\rangle+ 6\langle(L_1+L_2)^2L_3^2\rangle+ 4\langle(L_1+L_2)L_3^3\rangle+\langle(L_3)^4\rangle.
\end{align}
We will show that 1PR diagrams cancel away in each of the terms separately in the above expression.

\subsubsection{$\langle(L_1+L_2)^4\rangle$}

The first term in (\ref{3l1}) reads,
\begin{align}\label{3l11}
    \langle(L_1+L_2)^4\rangle = \langle L_1^4\rangle + 4\langle L_1^3L_2\rangle + 6\langle L_1^2L_2^2\rangle + 4\langle L_1L_2^3\rangle + \langle L_2^4\rangle
\end{align}
It is evident from the expression of $ L_2$ that $\langle L_2^4\rangle$ contains only disconnected diagrams which are expected to cancel with those coming from $\frac{i\hbar^3}{2}\Big\langle iA_4 - \frac{1}{2}A_3^2\Big\rangle^2$ in (\ref{3l}). In what follows we will focus only on connected 1PR diagrams thereby ignoring the last term in (\ref{3l11}). Since Wick's theorem leads to numerous terms if written down mathematically we represent them in the form of Feynman diagrams to reduce the clutter of expressions. 

The first term, $\langle L_1^4\rangle$, in (\ref{3l11}) reads the following in the diagrammatic form.

\begin{align}\label{3l11c1}
    &\dfrac{(3^4. 2) . ({4\choose 2}. 2)}{6^4}\begin{tikzpicture}[baseline={([yshift=-.5ex]current bounding box.center)}]
    \begin{feynman}
    \vertex [dot] (a) at (-1.5,0) ;
    \node [dot] (b) at (-1,0);
    \node [dot] (c) at (-0.5,0);
    \node [dot] (d) at (0,0);
    \node [dot] (e) at (0.5,0);
    \vertex  (f) at (1,0);
    \diagram*{
    (a) -- [red, half right] (b) -- [red] (c) -- [red, half right] (d) -- [red] (e) -- [red, half right] (f) -- [red, half right] (e) -- [red] (d) -- [red, half right] (c) -- [red] (b) -- [red, half right] (a)
    };
    \end{feynman}
    \end{tikzpicture} 
    + 
    \frac{(3^4 . 2^2) . (4. 3)}{6^4}\begin{tikzpicture}[baseline={([yshift=-.5ex]current bounding box.center)}]
    \begin{feynman}
    \vertex (a) at (0,0);
    \node [dot] (b) at (0.5,0);
    \node [dot] (c) at (1,0);
    \node [dot] (d) at (1.35,0.35);
    \node [dot] (e) at (1.35,-0.35);
    \diagram*{ (a) -- [red, half right] (b) -- [red] (c) -- [red] (d) -- [red] (e) -- [red, half right] (d) -- [red] (e) -- [red] (c) -- [red] (b) -- [red, half right] (a) };
    \end{feynman}
    \end{tikzpicture}
    +
    \frac{(3^4. 2) . (4)}{6^4}\begin{tikzpicture}[baseline={([yshift=-.5ex]current bounding box.center)}]
    \begin{feynman}
    \vertex (a) at (0,0);
    \node [dot] (b) at (0.5,0);
    \node [dot] (c) at (1,0);
    \node [dot] (d) at (1.35,0.35);
    \node [dot] (e) at (1.35,-0.35);
    \vertex (f) at (1.7,0.7);
    \vertex (g) at (1.7,-0.7);
    \diagram*{(a) -- [red, half right] (b) -- [red] (c) -- [red] (d) -- [red, half right] (f) -- [red, half right] (d) -- [red] (c) -- [red] (e) -- [red, half right] (g) -- [red, half right] (e) -- [red] (c) -- [red] (b) -- [red, half right] (a)};
    \end{feynman}
    \end{tikzpicture}
\end{align}

 The numerical factors attached to the diagrams can be explained as follows. Exactly three lines emerge from all four vertices in all the diagrams in (\ref{3l11c1}). At a particular vertex, a line may join with another line from the same vertex forming a loop; or another line from a different vertex forming a propagator. Consider the first diagram in (\ref{3l11c1}).

\begin{align}\label{rule1}
  \dfrac{(3^4. 2) . ({4\choose 2}. 2)}{6^4} \begin{tikzpicture}[baseline={([yshift=-.5ex]current bounding box.center)}]
    \begin{feynman}
    \vertex  (a) at (-1.5,0) ;
    \node [dot,label={\small 1}] (b) at (-1,0);
    \node [dot,label={\small 2}] (c) at (-0.5,0);
    \node [dot,label={\small 3}] (d) at (0,0);
    \node [dot,label={\small 4}] (e) at (0.5,0);
    \vertex  (f) at (1,0);
    \diagram*{
    (a) -- [red, half right] (b) -- [red] (c) -- [red, half right] (d) -- [red] (e) -- [red, half right] (f) -- [red, half right] (e) -- [red] (d) -- [red, half right] (c) -- [red] (b) -- [red, half right] (a)
    };
    \end{feynman}
    \end{tikzpicture}
\end{align}

We see that
\begin{enumerate}
    \item there are 3 ways to choose a line from each vertex to form a propagator $\rightarrow$ $\times 3^4$,
    \item there are 2 ways to form a loop between vertex (2) and (3)$\rightarrow$ $\times 2$,
    \item and, finally there are ${4\choose 2} \times 2$ ways to swap the position of vertices$\rightarrow$ $\times{4\choose 2} \times 2$.
\end{enumerate}
The convention of writing the numbers before the diagrams is as follows. The product of numbers in the first bracket tells us the number of ways to join lines with a fixed configuration of vertices, and the product of numbers in the second bracket tells us the number of ways to swap the position of vertices. One can explain the numerical factors attached to other diagrams as well in the same way.

Following are the diagrammatic expressions for $4\langle L_1^3L_2\rangle$, $6\langle L_1^2L_2^2\rangle$, and $4\langle L_1L_2^3\rangle$.
\begin{align}\label{3l11c2}
    4\langle L_1^3L_2\rangle=-4\Big(\dfrac{(3^3. 2). (3. 2)}{6^4\times 2}\begin{tikzpicture}[baseline={([yshift=-.5ex]current bounding box.center)}]
    \begin{feynman}
    \vertex [dot] (a) at (-1.5,0) ;
    \node [dot] (b) at (-1,0);
    \node [dot] (c) at (-0.5,0);
    \node [dot] (d) at (0,0);
    \node [dot] (e) at (0.5,0);
    \vertex  (f) at (1,0);
    \diagram*{
    (a) -- [red, half right] (b) -- [red] (c) -- [red, half right] (d) -- [red] (e) -- [red, half right] (f) -- [red, half right] (e) -- [red] (d) -- [red, half right] (c) -- [red] (b) -- [red, half right] (a)
    };
    \end{feynman}
    \end{tikzpicture} 
    &+ 
    \frac{(3^3. 2^2). (3)}{6^3\times 2}\begin{tikzpicture}[baseline={([yshift=-.5ex]current bounding box.center)}]
    \begin{feynman}
    \vertex (a) at (0,0);
    \node [dot] (b) at (0.5,0);
    \node [dot] (c) at (1,0);
    \node [dot] (d) at (1.35,0.35);
    \node [dot] (e) at (1.35,-0.35);
    \diagram*{ (a) -- [red, half right] (b) -- [red] (c) -- [red] (d) -- [red] (e) -- [red, half right] (d) -- [red] (e) -- [red] (c) -- [red] (b) -- [red, half right] (a) };
    \end{feynman}
    \end{tikzpicture}
    \nonumber\\&+
    \frac{(3^3. 2). (3)}{6^3\times 2}\begin{tikzpicture}[baseline={([yshift=-.5ex]current bounding box.center)}]
    \begin{feynman}
    \vertex (a) at (0,0);
    \node [dot] (b) at (0.5,0);
    \node [dot] (c) at (1,0);
    \node [dot] (d) at (1.35,0.35);
    \node [dot] (e) at (1.35,-0.35);
    \vertex (f) at (1.7,0.7);
    \vertex (g) at (1.7,-0.7);
    \diagram*{(a) -- [red, half right] (b) -- [red] (c) -- [red] (d) -- [red, half right] (f) -- [red, half right] (d) -- [red] (c) -- [red] (e) -- [red, half right] (g) -- [red, half right] (e) -- [red] (c) -- [red] (b) -- [red, half right] (a)};
    \end{feynman}
    \end{tikzpicture}\Big)
\end{align}

\begin{align}\label{3l11c3}
    6\langle L_1^2L_2^2\rangle=6\Big(\dfrac{(3^2. 2). (2)}{6^2\times 2^2}\begin{tikzpicture}[baseline={([yshift=-.5ex]current bounding box.center)}]
    \begin{feynman}
    \vertex [dot] (a) at (-1.5,0) ;
    \node [dot] (b) at (-1,0);
    \node [dot] (c) at (-0.5,0);
    \node [dot] (d) at (0,0);
    \node [dot] (e) at (0.5,0);
    \vertex  (f) at (1,0);
    \diagram*{
    (a) -- [red, half right] (b) -- [red] (c) -- [red, half right] (d) -- [red] (e) -- [red, half right] (f) -- [red, half right] (e) -- [red] (d) -- [red, half right] (c) -- [red] (b) -- [red, half right] (a)
    };
    \end{feynman}
    \end{tikzpicture} 
    +
    \frac{(3^2. 2). (2)}{6^2\times 2^2}\begin{tikzpicture}[baseline={([yshift=-.5ex]current bounding box.center)}]
    \begin{feynman}
    \vertex (a) at (0,0);
    \node [dot] (b) at (0.5,0);
    \node [dot] (c) at (1,0);
    \node [dot] (d) at (1.35,0.35);
    \node [dot] (e) at (1.35,-0.35);
    \vertex (f) at (1.7,0.7);
    \vertex (g) at (1.7,-0.7);
    \diagram*{(a) -- [red, half right] (b) -- [red] (c) -- [red] (d) -- [red, half right] (f) -- [red, half right] (d) -- [red] (c) -- [red] (e) -- [red, half right] (g) -- [red, half right] (e) -- [red] (c) -- [red] (b) -- [red, half right] (a)};
    \end{feynman}
    \end{tikzpicture}\Big)
\end{align}

\begin{align}\label{3l11c4}
    4\langle L_1L_2^3\rangle=-4\Big(\dfrac{(3). (2)}{6\times 2^3}\begin{tikzpicture}[baseline={([yshift=-.5ex]current bounding box.center)}]
    \begin{feynman}
    \vertex (a) at (0,0);
    \node [dot] (b) at (0.5,0);
    \node [dot] (c) at (1,0);
    \node [dot] (d) at (1.35,0.35);
    \node [dot] (e) at (1.35,-0.35);
    \vertex (f) at (1.7,0.7);
    \vertex (g) at (1.7,-0.7);
    \diagram*{(a) -- [red, half right] (b) -- [red] (c) -- [red] (d) -- [red, half right] (f) -- [red, half right] (d) -- [red] (c) -- [red] (e) -- [red, half right] (g) -- [red, half right] (e) -- [red] (c) -- [red] (b) -- [red, half right] (a)};
    \end{feynman}
    \end{tikzpicture}\Big)
\end{align}
On adding (\ref{3l11c1}), (\ref{3l11c2}), (\ref{3l11c3}) and (\ref{3l11c4}) we see that all 1PR diagrams cancel away.

\subsubsection{$\langle(L_1+L_2)^3L_3\rangle$}

To show that the remaining four terms in (\ref{3l1}) do not contain any 1PR diagrams we need to use (\descref{Result 1}) and (\descref{Result 2}) obtained in Section (\ref{SecV}). Consider the second term in (\ref{3l1}).
\begin{align}\label{3l2}
    \langle(L_1+L_2)^3L_3\rangle = \langle(L_1^3L_3 + 3L_1^2L_2L_3 + 3L_1L_2^2L_3 + L_2^3L_3\rangle
\end{align}
It contains only one vertex with $S_{;ij}$ (the $L_3$ term). The resulting diagrams can be divided into two mutually exclusive categories.

\begin{description}
   \item[Category 1:] The diagrams in which all the lines coming out of the vertex with $S_{;ij}$ join with lines coming from a different vertex.
   \item[Category 2:] The diagrams in which a pair of lines coming out of vertex with $S_{;ij}$ join with each other forming a loop.
\end{description}
    
Let us focus on the diagrams belonging to category (1). We find the following diagrammatic expression 
corresponding to each term in (\ref{3l2}).
\begin{align}\label{3l2c11}
   &\langle(L_1^3L_3)\rangle \rightarrow \frac{1}{6^3}\Big((3^4. 2). (3. 2)\begin{tikzpicture}[baseline={([yshift=-.5ex]current bounding box.center)}]
    \begin{feynman}
    \vertex [dot] (a) at (-1.5,0) ;
    \node [dot] (b) at (-1,0);
    \node [empty dot, fill=blue] (c) at (-0.5,0);
    \node [dot] (d) at (0,0);
    \node [dot] (e) at (0.5,0);
    \vertex  (f) at (1,0);
    \diagram*{
    (a) -- [red, half right] (b) -- [red] (c) -- [red, half right] (d) -- [red] (e) -- [red, half right] (f) -- [red, half right] (e) -- [red] (d) -- [red, half right] (c) -- [red] (b) -- [red, half right] (a)
    };
    \end{feynman}
    \end{tikzpicture} + (3^4. 2^2). (3. 3)\begin{tikzpicture}[baseline={([yshift=-.5ex]current bounding box.center)}]
    \begin{feynman}
    \vertex (a) at (0,0);
    \node [dot] (b) at (0.5,0);
    \node [dot] (c) at (1,0);
    \node [empty dot, fill=blue] (d) at (1.35,0.35);
    \node [dot] (e) at (1.35,-0.35);
    \diagram*{ (a) -- [red, half right] (b) -- [red] (c) -- [red] (d) -- [red] (e) -- [red, half right] (d) -- [red] (e) -- [red] (c) -- [red] (b) -- [red, half right] (a) };
    \end{feynman}
    \end{tikzpicture} \nonumber\\
    &\hspace{90mm}+(3^4.2).(1)\begin{tikzpicture}[baseline={([yshift=-.5ex]current bounding box.center)}]
    \begin{feynman}
    \vertex (a) at (0,0);
    \node [dot] (b) at (0.5,0);
    \node [empty dot, fill=blue] (c) at (1,0);
    \node [dot] (d) at (1.35,0.35);
    \node [dot] (e) at (1.35,-0.35);
    \vertex (f) at (1.7,0.7);
    \vertex (g) at (1.7,-0.7);
    \diagram*{(a) -- [red, half right] (b) -- [red] (c) -- [red] (d) -- [red, half right] (f) -- [red, half right] (d) -- [red] (c) -- [red] (e) -- [red, half right] (g) -- [red, half right] (e) -- [red] (c) -- [red] (b) -- [red, half right] (a)};
    \end{feynman}
    \end{tikzpicture}\Big),\\
    &3\langle(L_1^2L_2L_3)\rangle \rightarrow -\frac{3}{6^2.2}\Big((3^3. 2).(4).\begin{tikzpicture}[baseline={([yshift=-.5ex]current bounding box.center)}]
    \begin{feynman}
    \vertex  (a) at (-1.5,0)  ;
    \node [dot] (b) at (-1,0);
    \node [empty dot, fill=blue] (c) at (-0.5,0);
    \node [dot] (d) at (0,0);
    \node [dot] (e) at (0.5,0);
    \vertex  (f) at (1,0);
    \diagram*{
    (a) -- [red, half right] (b) -- [red] (c) -- [red, half right] (d) -- [red] (e) -- [red, half right] (f) -- [red, half right] (e) -- [red] (d) -- [red, half right] (c) -- [red] (b) -- [red, half right] (a)
    };
    \end{feynman}
    \end{tikzpicture}+(3^3.2^2).(3).\begin{tikzpicture}[baseline={([yshift=-.5ex]current bounding box.center)}]
    \begin{feynman}
    \vertex (a) at (0,0);
    \node [dot] (b) at (0.5,0);
    \node [dot] (c) at (1,0);
    \node [empty dot, fill=blue] (d) at (1.35,0.35);
    \node [dot] (e) at (1.35,-0.35);
    \diagram*{ (a) -- [red, half right] (b) -- [red] (c) -- [red] (d) -- [red] (e) -- [red, half right] (d) -- [red] (e) -- [red] (c) -- [red] (b) -- [red, half right] (a) };
    \end{feynman}
    \end{tikzpicture}\nonumber\\
    &\hspace{90mm}+(3^3.2).(1).\begin{tikzpicture}[baseline={([yshift=-.5ex]current bounding box.center)}]
    \begin{feynman}
    \vertex (a) at (0,0);
    \node [dot] (b) at (0.5,0);
    \node [empty dot, fill=blue] (c) at (1,0);
    \node [dot] (d) at (1.35,0.35);
    \node [dot] (e) at (1.35,-0.35);
    \vertex (f) at (1.7,0.7);
    \vertex (g) at (1.7,-0.7);
    \diagram*{(a) -- [red, half right] (b) -- [red] (c) -- [red] (d) -- [red, half right] (f) -- [red, half right] (d) -- [red] (c) -- [red] (e) -- [red, half right] (g) -- [red, half right] (e) -- [red] (c) -- [red] (b) -- [red, half right] (a)};
    \end{feynman}
    \end{tikzpicture}\Big),\label{3l2c12}\\
    &3\langle(L_1L_2^2L_3)\rangle \rightarrow \frac{3}{6.2^2}\Big((3^2. 2). (2).\begin{tikzpicture}[baseline={([yshift=-.5ex]current bounding box.center)}]
    \begin{feynman}
    \vertex [dot] (a) at (-1.5,0) ;
    \node [dot] (b) at (-1,0);
    \node [empty dot, fill=blue] (c) at (-0.5,0);
    \node [dot] (d) at (0,0);
    \node [dot] (e) at (0.5,0);
    \vertex  (f) at (1,0);
    \diagram*{
    (a) -- [red, half right] (b) -- [red] (c) -- [red, half right] (d) -- [red] (e) -- [red, half right] (f) -- [red, half right] (e) -- [red] (d) -- [red, half right] (c) -- [red] (b) -- [red, half right] (a)
    };
    \end{feynman}
    \end{tikzpicture} +(3^2.2).(1).\begin{tikzpicture}[baseline={([yshift=-.5ex]current bounding box.center)}]
    \begin{feynman}
    \vertex (a) at (0,0);
    \node [dot] (b) at (0.5,0);
    \node [empty dot, fill=blue] (c) at (1,0);
    \node [dot] (d) at (1.35,0.35);
    \node [dot] (e) at (1.35,-0.35);
    \vertex (f) at (1.7,0.7);
    \vertex (g) at (1.7,-0.7);
    \diagram*{(a) -- [red, half right] (b) -- [red] (c) -- [red] (d) -- [red, half right] (f) -- [red, half right] (d) -- [red] (c) -- [red] (e) -- [red, half right] (g) -- [red, half right] (e) -- [red] (c) -- [red] (b) -- [red, half right] (a)};
    \end{feynman}
    \end{tikzpicture}\Big),\label{3l2c13}\\
    &\langle L_2^3L_3\rangle \rightarrow -\frac{(3.2).(1)}{2^3}.\begin{tikzpicture}[baseline={([yshift=-.5ex]current bounding box.center)}]
    \begin{feynman}
    \vertex (a) at (0,0);
    \node [dot] (b) at (0.5,0);
    \node [empty dot, fill=blue] (c) at (1,0);
    \node [dot] (d) at (1.35,0.35);
    \node [dot] (e) at (1.35,-0.35);
    \vertex (f) at (1.7,0.7);
    \vertex (g) at (1.7,-0.7);
    \diagram*{(a) -- [red, half right] (b) -- [red] (c) -- [red] (d) -- [red, half right] (f) -- [red, half right] (d) -- [red] (c) -- [red] (e) -- [red, half right] (g) -- [red, half right] (e) -- [red] (c) -- [red] (b) -- [red, half right] (a)};
    \end{feynman}
    \end{tikzpicture}.\label{3l2c14}
\end{align}
The blue dot represents the vertex with $S_{;ij}$ in the diagrams above. The numbers appearing in front of the diagrams are obtained in the same way as for (\ref{rule1}) with the additional rule that the diagrams must belong to category (1). On adding (\ref{3l2c11}), (\ref{3l2c12}), (\ref{3l2c13}) and (\ref{3l2c14}) we find that all 1PR diagrams of category (1) cancel away.

Now let us look at the diagrams of category (2). The diagrams in this category are such that the lines coming out from the vertex in $L_3$ form a loop. There are three ways to do this; 
\begin{enumerate}
    \item $S_{;ij}\Gamma^j{}_{kl}\langle\eta^i\eta^k\rangle\eta^l \propto \Gamma^j{}_{jl}\propto\delta^4(0)$
    \item $S_{;ij}\Gamma^j{}_{kl}\langle\eta^i\eta^l\rangle\eta^k \propto \Gamma^j{}_{jk}\propto\delta^4(0)$
    \item $S_{;ij}\Gamma^j{}_{kl}\langle\eta^k\eta^l\rangle\eta^i$
\end{enumerate}
Out of these the first two are regularised to zero following (\descref{Result 1}) found in Section (\ref{SecV}). Therefore, only the last term can contribute to the diagrams in category (2). We list all these diagrams with the correct numerical coefficients as per the rule stated above for all the terms in (\ref{3l2}).
\begin{align}\label{3l2c21}
   &\langle(L_1^3L_3)\rangle \rightarrow \frac{1}{6^3}\Big((3^3. 2). (3. 2).\begin{tikzpicture}[baseline={([yshift=-.5ex]current bounding box.center)}]
    \begin{feynman}
    \vertex [dot] (a) at (-1.5,0) ;
    \node [dot] (b) at (-1,0);
    \node [dot] (c) at (-0.5,0);
    \node [dot] (d) at (0,0);
    \node [empty dot, fill=blue] (e) at (0.5,0);
    \vertex  (f) at (1,0);
    \diagram*{
    (a) -- [red, half right] (b) -- [red] (c) -- [red, half right] (d) -- [red] (e) -- [red, half right] (f) -- [red, half right] (e) -- [red] (d) -- [red, half right] (c) -- [red] (b) -- [red, half right] (a)
    };
    \end{feynman}
    \end{tikzpicture} + (3^3.2).(3).\begin{tikzpicture}[baseline={([yshift=-.5ex]current bounding box.center)}]
    \begin{feynman}
    \vertex (a) at (0,0);
    \node [dot] (b) at (0.5,0);
    \node [dot] (c) at (1,0);
    \node [empty dot, fill=blue] (d) at (1.35,0.35);
    \node [dot] (e) at (1.35,-0.35);
    \vertex (f) at (1.7,0.7);
    \vertex (g) at (1.7,-0.7);
    \diagram*{(a) -- [red, half right] (b) -- [red] (c) -- [red] (d) -- [red, half right] (f) -- [red, half right] (d) -- [red] (c) -- [red] (e) -- [red, half right] (g) -- [red, half right] (e) -- [red] (c) -- [red] (b) -- [red, half right] (a)};
    \end{feynman}
    \end{tikzpicture}\Big)\\
    &3\langle(L_1^2L_2L_3)\rangle \rightarrow -\frac{3}{6^2.2}\Big((3^2. 2).(2).\begin{tikzpicture}[baseline={([yshift=-.5ex]current bounding box.center)}]
    \begin{feynman}
    \vertex  (a) at (-1.5,0)  ;
    \node [dot] (b) at (-1,0);
    \node [dot] (c) at (-0.5,0);
    \node [dot] (d) at (0,0);
    \node [empty dot, fill=blue] (e) at (0.5,0);
    \vertex  (f) at (1,0);
    \diagram*{
    (a) -- [red, half right] (b) -- [red] (c) -- [red, half right] (d) -- [red] (e) -- [red, half right] (f) -- [red, half right] (e) -- [red] (d) -- [red, half right] (c) -- [red] (b) -- [red, half right] (a)
    };
    \end{feynman}
    \end{tikzpicture}+(3^2.2).(2).\begin{tikzpicture}[baseline={([yshift=-.5ex]current bounding box.center)}]
    \begin{feynman}
    \vertex (a) at (0,0);
    \node [dot] (b) at (0.5,0);
    \node [dot] (c) at (1,0);
    \node [empty dot, fill=blue] (d) at (1.35,0.35);
    \node [dot] (e) at (1.35,-0.35);
    \vertex (f) at (1.7,0.7);
    \vertex (g) at (1.7,-0.7);
    \diagram*{(a) -- [red, half right] (b) -- [red] (c) -- [red] (d) -- [red, half right] (f) -- [red, half right] (d) -- [red] (c) -- [red] (e) -- [red, half right] (g) -- [red, half right] (e) -- [red] (c) -- [red] (b) -- [red, half right] (a)};
    \end{feynman}
    \end{tikzpicture}\Big)\label{3l2c22}\\
    &3\langle(L_1L_2^2L_3)\rangle \rightarrow \frac{3}{6.2^2}.(3.2).(1).\begin{tikzpicture}[baseline={([yshift=-.5ex]current bounding box.center)}]
    \begin{feynman}
    \vertex (a) at (0,0);
    \node [dot] (b) at (0.5,0);
    \node [dot] (c) at (1,0);
    \node [empty dot, fill=blue] (d) at (1.35,0.35);
    \node [dot] (e) at (1.35,-0.35);
    \vertex (f) at (1.7,0.7);
    \vertex (g) at (1.7,-0.7);
    \diagram*{(a) -- [red, half right] (b) -- [red] (c) -- [red] (d) -- [red, half right] (f) -- [red, half right] (d) -- [red] (c) -- [red] (e) -- [red, half right] (g) -- [red, half right] (e) -- [red] (c) -- [red] (b) -- [red, half right] (a)};
    \end{feynman}
    \end{tikzpicture}\label{3l2c23}
\end{align}
We ignored the diagram, \begin{tikzpicture}[baseline={([yshift=-.5ex]current bounding box.center)}]
    \begin{feynman}
    \vertex (a) at (0,0);
    \node [empty dot, fill=blue] (b) at (0.5,0);
    \node [dot] (c) at (1,0);
    \node [dot] (d) at (1.35,0.35);
    \node [dot] (e) at (1.35,-0.35);
    \diagram*{ (a) -- [red, half right] (b) -- [red] (c) -- [red] (d) -- [red] (e) -- [red, half right] (d) -- [red] (e) -- [red] (c) -- [red] (b) -- [red, half right] (a) };
    \end{feynman}
    \end{tikzpicture}, in category (2) because either it is regularised to zero or it turns into a 1PI diagram as explained in (\descref{Result 1}) and (\descref{Result 2}). These results can also be used in (\ref{3l2c21}), (\ref{3l2c22}), and (\ref{3l2c23}) but these do not turn into 1PI diagrams. The ones that regularise to zero have been removed which is evident from the numerical factors (which are less by a factor of $3$ for each blue vertex) accompanying the diagrams with blue vertices. There is no need to use (\descref{Result 2}) for further simplification because on adding (\ref{3l2c21}), (\ref{3l2c22}), and (\ref{3l2c23}) we find that all 1PR diagrams of category (2) cancel away. Also, the last term, $\langle L_2^3L_3\rangle$, is ignored because it cannot contain any 1PR diagrams of category (2).

\subsubsection{$\langle(L_1+L_2)^2L_3^2\rangle$}

The third term reads
\begin{align}\label{3l13}
    \langle(L_1+L_2)^2L_3^2\rangle = \langle L_1^2L_3^2\rangle + 2\langle L_1L_2L_3^2\rangle + \langle L_2^2L_3^2\rangle.
\end{align}
We now have two vertices with $S_{;ij}$ and as such the diagrams in (\ref{3l13}) can be divided into three mutually exclusive categories based on how the lines from the vertices with $S_{;ij}$ join.

\begin{description}
   \item[Category 3:] The diagrams in which lines from both the vertices join as per category (1).
   \item[Category 4:] The diagrams in which lines from both the vertices join as per category (2).
   \item[Category 5:] The diagrams in which lines from one vertex join as per category (1) and lines from another vertex join as per category (2).
\end{description}

The diagrams arising from each term in (\ref{3l13}) of category (3) are listed below.

\begin{align}\label{3l13c31}
   &\langle(L_1^2L_3^2)\rangle = \frac{1}{6^2}\Big((3^4.2).(2).\begin{tikzpicture}[baseline={([yshift=-.5ex]current bounding box.center)}]
    \begin{feynman}
    \vertex [dot] (a) at (-1.5,0) ;
    \node [dot] (b) at (-1,0);
    \node [empty dot, fill=blue] (c) at (-0.5,0);
    \node [empty dot, fill=blue] (d) at (0,0);
    \node [dot] (e) at (0.5,0);
    \vertex  (f) at (1,0);
    \diagram*{
    (a) -- [red, half right] (b) -- [red] (c) -- [red, half right] (d) -- [red] (e) -- [red, half right] (f) -- [red, half right] (e) -- [red] (d) -- [red, half right] (c) -- [red] (b) -- [red, half right] (a)
    };
    \end{feynman}
    \end{tikzpicture}+(3^4.2^2).(2.3).\begin{tikzpicture}[baseline={([yshift=-.5ex]current bounding box.center)}]
    \begin{feynman}
    \vertex (a) at (0,0);
    \node [dot] (b) at (0.5,0);
    \node [dot] (c) at (1,0);
    \node [empty dot, fill=blue] (d) at (1.35,0.35);
    \node [empty dot, fill=blue] (e) at (1.35,-0.35);
    \diagram*{ (a) -- [red, half right] (b) -- [red] (c) -- [red] (d) -- [red] (e) -- [red, half right] (d) -- [red] (e) -- [red] (c) -- [red] (b) -- [red, half right] (a) };
    \end{feynman}
    \end{tikzpicture}\Big),\\
    &2\langle(L_1L_2L_3^2)\rangle = -\frac{2}{6.2}\Big((3^3.2).(2).\begin{tikzpicture}[baseline={([yshift=-.5ex]current bounding box.center)}]
    \begin{feynman}
    \vertex [dot] (a) at (-1.5,0) ;
    \node [dot] (b) at (-1,0);
    \node [empty dot, fill=blue] (c) at (-0.5,0);
    \node [empty dot, fill=blue] (d) at (0,0);
    \node [dot] (e) at (0.5,0);
    \vertex  (f) at (1,0);
    \diagram*{
    (a) -- [red, half right] (b) -- [red] (c) -- [red, half right] (d) -- [red] (e) -- [red, half right] (f) -- [red, half right] (e) -- [red] (d) -- [red, half right] (c) -- [red] (b) -- [red, half right] (a)
    };
    \end{feynman}
    \end{tikzpicture}+(3^3.2^2).(1).\begin{tikzpicture}[baseline={([yshift=-.5ex]current bounding box.center)}]
    \begin{feynman}
    \vertex (a) at (0,0);
    \node [dot] (b) at (0.5,0);
    \node [dot] (c) at (1,0);
    \node [empty dot, fill=blue] (d) at (1.35,0.35);
    \node [empty dot, fill=blue] (e) at (1.35,-0.35);
    \diagram*{ (a) -- [red, half right] (b) -- [red] (c) -- [red] (d) -- [red] (e) -- [red, half right] (d) -- [red] (e) -- [red] (c) -- [red] (b) -- [red, half right] (a) };
    \end{feynman}
    \end{tikzpicture}\Big),\label{3l13c32}\\
    &\langle(L_2^2L_3^2)\rangle = \frac{1}{2^2}(3^2.2).(2).\begin{tikzpicture}[baseline={([yshift=-.5ex]current bounding box.center)}]
    \begin{feynman}
    \vertex [dot] (a) at (-1.5,0) ;
    \node [dot] (b) at (-1,0);
    \node [empty dot, fill=blue] (c) at (-0.5,0);
    \node [empty dot, fill=blue] (d) at (0,0);
    \node [dot] (e) at (0.5,0);
    \vertex  (f) at (1,0);
    \diagram*{
    (a) -- [red, half right] (b) -- [red] (c) -- [red, half right] (d) -- [red] (e) -- [red, half right] (f) -- [red, half right] (e) -- [red] (d) -- [red, half right] (c) -- [red] (b) -- [red, half right] (a)
    };
    \end{feynman}
    \end{tikzpicture}.
\end{align}

The diagrams arising from each term in (\ref{3l13}) of category (4) are

\begin{align}\label{3l13c41}
   &\langle(L_1^2L_3^2)\rangle \rightarrow \frac{1}{6^2}.(3^2.2).(2).\begin{tikzpicture}[baseline={([yshift=-.5ex]current bounding box.center)}]
    \begin{feynman}
    \vertex (a) at (0,0);
    \node [dot] (b) at (0.5,0);
    \node [dot] (c) at (1,0);
    \node [empty dot, fill=blue] (d) at (1.35,0.35);
    \node [empty dot, fill=blue] (e) at (1.35,-0.35);
    \vertex (f) at (1.7,0.7);
    \vertex (g) at (1.7,-0.7);
    \diagram*{(a) -- [red, half right] (b) -- [red] (c) -- [red] (d) -- [red, half right] (f) -- [red, half right] (d) -- [red] (c) -- [red] (e) -- [red, half right] (g) -- [red, half right] (e) -- [red] (c) -- [red] (b) -- [red, half right] (a)};
    \end{feynman}
    \end{tikzpicture},\\
    &2\langle(L_1L_2L_3^2)\rangle \rightarrow -\frac{2}{6.2}(3.2).(1).\begin{tikzpicture}[baseline={([yshift=-.5ex]current bounding box.center)}]
    \begin{feynman}
    \vertex (a) at (0,0);
    \node [dot] (b) at (0.5,0);
    \node [dot] (c) at (1,0);
    \node [empty dot, fill=blue] (d) at (1.35,0.35);
    \node [empty dot, fill=blue] (e) at (1.35,-0.35);
    \vertex (f) at (1.7,0.7);
    \vertex (g) at (1.7,-0.7);
    \diagram*{(a) -- [red, half right] (b) -- [red] (c) -- [red] (d) -- [red, half right] (f) -- [red, half right] (d) -- [red] (c) -- [red] (e) -- [red, half right] (g) -- [red, half right] (e) -- [red] (c) -- [red] (b) -- [red, half right] (a)};
    \end{feynman}
    \end{tikzpicture}.\label{3l13c42}
\end{align}

and the diagrams arising from each term in (\ref{3l13}) of category (5) are

\begin{align}\label{3l13c51}
   &\langle(L_1^2L_3^2)\rangle \rightarrow \frac{1}{6^2}\Big((3^3.2).(2. 2.2).\begin{tikzpicture}[baseline={([yshift=-.5ex]current bounding box.center)}]
    \begin{feynman}
    \vertex [dot] (a) at (-1.5,0) ;
    \node [empty dot, fill=blue] (b) at (-1,0);
    \node [empty dot, fill=blue] (c) at (-0.5,0);
    \node [dot] (d) at (0,0);
    \node [dot] (e) at (0.5,0);
    \vertex  (f) at (1,0);
    \diagram*{
    (a) -- [red, half right] (b) -- [red] (c) -- [red, half right] (d) -- [red] (e) -- [red, half right] (f) -- [red, half right] (e) -- [red] (d) -- [red, half right] (c) -- [red] (b) -- [red, half right] (a)
    };
    \end{feynman}
    \end{tikzpicture} + (3^3.2).(2).\begin{tikzpicture}[baseline={([yshift=-.5ex]current bounding box.center)}]
    \begin{feynman}
    \vertex (a) at (0,0);
    \node [empty dot, fill=blue] (b) at (0.5,0);
    \node [empty dot, fill=blue] (c) at (1,0);
    \node [dot] (d) at (1.35,0.35);
    \node [dot] (e) at (1.35,-0.35);
    \vertex (f) at (1.7,0.7);
    \vertex (g) at (1.7,-0.7);
    \diagram*{(a) -- [red, half right] (b) -- [red] (c) -- [red] (d) -- [red, half right] (f) -- [red, half right] (d) -- [red] (c) -- [red] (e) -- [red, half right] (g) -- [red, half right] (e) -- [red] (c) -- [red] (b) -- [red, half right] (a)};
    \end{feynman}
    \end{tikzpicture}\Big),\\
    &2\langle(L_1L_2L_2L_3^2)\rangle \rightarrow -\frac{2}{6.2}\Big((3^2. 2).(2.2).\begin{tikzpicture}[baseline={([yshift=-.5ex]current bounding box.center)}]
    \begin{feynman}
    \vertex  (a) at (-1.5,0)  ;
    \node [empty dot, fill=blue] (b) at (-1,0);
    \node [empty dot, fill=blue] (c) at (-0.5,0);
    \node [dot] (d) at (0,0);
    \node [dot] (e) at (0.5,0);
    \vertex  (f) at (1,0);
    \diagram*{
    (a) -- [red, half right] (b) -- [red] (c) -- [red, half right] (d) -- [red] (e) -- [red, half right] (f) -- [red, half right] (e) -- [red] (d) -- [red, half right] (c) -- [red] (b) -- [red, half right] (a)
    };
    \end{feynman}
    \end{tikzpicture}+(3^2.2).(2).\begin{tikzpicture}[baseline={([yshift=-.5ex]current bounding box.center)}]
    \begin{feynman}
    \vertex (a) at (0,0);
    \node [empty dot, fill=blue] (b) at (0.5,0);
    \node [empty dot, fill=blue] (c) at (1,0);
    \node [dot] (d) at (1.35,0.35);
    \node [dot] (e) at (1.35,-0.35);
    \vertex (f) at (1.7,0.7);
    \vertex (g) at (1.7,-0.7);
    \diagram*{(a) -- [red, half right] (b) -- [red] (c) -- [red] (d) -- [red, half right] (f) -- [red, half right] (d) -- [red] (c) -- [red] (e) -- [red, half right] (g) -- [red, half right] (e) -- [red] (c) -- [red] (b) -- [red, half right] (a)};
    \end{feynman}
    \end{tikzpicture}\Big),\label{3l13c52}\\
    &\langle(L_2^2L_3^2)\rangle \rightarrow \frac{1}{2^2}(3.2).(2).\begin{tikzpicture}[baseline={([yshift=-.5ex]current bounding box.center)}]
    \begin{feynman}
    \vertex (a) at (0,0);
    \node [empty dot, fill=blue] (b) at (0.5,0);
    \node [empty dot, fill=blue] (c) at (1,0);
    \node [dot] (d) at (1.35,0.35);
    \node [dot] (e) at (1.35,-0.35);
    \vertex (f) at (1.7,0.7);
    \vertex (g) at (1.7,-0.7);
    \diagram*{(a) -- [red, half right] (b) -- [red] (c) -- [red] (d) -- [red, half right] (f) -- [red, half right] (d) -- [red] (c) -- [red] (e) -- [red, half right] (g) -- [red, half right] (e) -- [red] (c) -- [red] (b) -- [red, half right] (a)};
    \end{feynman}
    \end{tikzpicture}.\label{3l13c53}
\end{align}
Again the absence of \begin{tikzpicture}[baseline={([yshift=-.5ex]current bounding box.center)}]
    \begin{feynman}
    \vertex [dot] (a) at (-1.5,0) ;
    \node [empty dot, fill=blue] (b) at (-1,0);
    \node [dot] (c) at (-0.5,0);
    \node [dot] (d) at (0,0);
    \node [empty dot, fill=blue] (e) at (0.5,0);
    \vertex  (f) at (1,0);
    \diagram*{
    (a) -- [red, half right] (b) -- [red] (c) -- [red, half right] (d) -- [red] (e) -- [red, half right] (f) -- [red, half right] (e) -- [red] (d) -- [red, half right] (c) -- [red] (b) -- [red, half right] (a)
    };
    \end{feynman}
    \end{tikzpicture} in category (4) and \begin{tikzpicture}[baseline={([yshift=-.5ex]current bounding box.center)}]
    \begin{feynman}
    \vertex (a) at (0,0);
    \node [empty dot, fill=blue] (b) at (0.5,0);
    \node [empty dot, fill=blue] (c) at (1,0);
    \node [dot] (d) at (1.35,0.35);
    \node [dot] (e) at (1.35,-0.35);
    \diagram*{ (a) -- [red, half right] (b) -- [red] (c) -- [red] (d) -- [red] (e) -- [red, half right] (d) -- [red] (e) -- [red] (c) -- [red] (b) -- [red, half right] (a) };
    \end{feynman}
    \end{tikzpicture} in category (5) can be explained based on (\descref{Result 1}) and (\descref{Result 2}) derived earlier. We can see that all 1PR diagrams in each category cancel away separately.

\subsubsection{$\langle(L_1+L_2)L_3^3\rangle$}

The fourth term in (\ref{3l1}) contains three vertices with $S_{;ij}$.
\begin{align}\label{3l14}
    \langle(L_1+L_2)L_3^3\rangle = \langle L_1L_3^3\rangle + \langle L_2L_3^3\rangle.
\end{align}

We can form four mutually exclusive categories into which the diagrams arising from (\ref{3l14}) can be sorted based on how the lines coming out from the vertices with $S_{;ij}$ join.
\begin{description}
   \item[Category 6:] The diagrams in which the lines from all three vertices join as per category (1).
   \item[Category 7:] The diagrams in which the lines from two of the three vertices join as per category (1) and lines from the remaining one join as per category (2).
   \item[Category 8:] The diagrams in which the lines from two of the three vertices join as per category (2) and the lines from the remaining one join as per category (1).
   \item[Category 9:] The diagrams in which the lines from all three vertices join as per category (2).
\end{description}

The diagrams arising from the terms in (\ref{3l14}) in category (6) are

\begin{align}\label{3l14c61}
   &\langle(L_1L_3^3)\rangle \rightarrow \frac{1}{6}.(3^4.2^2).(3).\begin{tikzpicture}[baseline={([yshift=-.5ex]current bounding box.center)}]
    \begin{feynman}
    \vertex (a) at (0,0);
    \node [dot] (b) at (0.5,0);
    \node [empty dot, fill=blue] (c) at (1,0);
    \node [empty dot, fill=blue] (d) at (1.35,0.35);
    \node [empty dot, fill=blue] (e) at (1.35,-0.35);
    \diagram*{ (a) -- [red, half right] (b) -- [red] (c) -- [red] (d) -- [red] (e) -- [red, half right] (d) -- [red] (e) -- [red] (c) -- [red] (b) -- [red, half right] (a) };
    \end{feynman}
    \end{tikzpicture},\\
    &\langle(L_2L_3^3)\rangle \rightarrow -\frac{1}{2}(3^3.2^2).(3).\begin{tikzpicture}[baseline={([yshift=-.5ex]current bounding box.center)}]
    \begin{feynman}
    \vertex (a) at (0,0);
    \node [dot] (b) at (0.5,0);
    \node [empty dot, fill=blue] (c) at (1,0);
    \node [empty dot, fill=blue] (d) at (1.35,0.35);
    \node [empty dot, fill=blue] (e) at (1.35,-0.35);
    \diagram*{ (a) -- [red, half right] (b) -- [red] (c) -- [red] (d) -- [red] (e) -- [red, half right] (d) -- [red] (e) -- [red] (c) -- [red] (b) -- [red, half right] (a) };
    \end{feynman}
    \end{tikzpicture}.\label{3l14c62}
\end{align}

The diagrams arising from the terms in (\ref{3l14}) in category (7) are

\begin{align}\label{3l14c71}
   &\langle(L_1L_3^3)\rangle \rightarrow \frac{1}{6}.(3^3.2).(3.2).\begin{tikzpicture}[baseline={([yshift=-.5ex]current bounding box.center)}]
    \begin{feynman}
    \vertex [dot] (a) at (-1.5,0) ;
    \node [empty dot, fill=blue] (b) at (-1,0);
    \node [empty dot, fill=blue] (c) at (-0.5,0);
    \node [empty dot, fill=blue] (d) at (0,0);
    \node [dot] (e) at (0.5,0);
    \vertex  (f) at (1,0);
    \diagram*{
    (a) -- [red, half right] (b) -- [red] (c) -- [red, half right] (d) -- [red] (e) -- [red, half right] (f) -- [red, half right] (e) -- [red] (d) -- [red, half right] (c) -- [red] (b) -- [red, half right] (a)
    };
    \end{feynman}
    \end{tikzpicture},\\
    &\langle(L_2L_3^3)\rangle \rightarrow -\frac{1}{2}(3^2.2).(3.2).\begin{tikzpicture}[baseline={([yshift=-.5ex]current bounding box.center)}]
    \begin{feynman}
    \vertex [dot] (a) at (-1.5,0) ;
    \node [empty dot, fill=blue] (b) at (-1,0);
    \node [empty dot, fill=blue] (c) at (-0.5,0);
    \node [empty dot, fill=blue] (d) at (0,0);
    \node [dot] (e) at (0.5,0);
    \vertex  (f) at (1,0);
    \diagram*{
    (a) -- [red, half right] (b) -- [red] (c) -- [red, half right] (d) -- [red] (e) -- [red, half right] (f) -- [red, half right] (e) -- [red] (d) -- [red, half right] (c) -- [red] (b) -- [red, half right] (a)
    };
    \end{feynman}
    \end{tikzpicture}.\label{3l14c72}
\end{align}

The diagrams arising from the terms in (\ref{3l14}) in category (8) are

\begin{align}\label{3l14c81}
   &\langle(L_1L_3^3)\rangle \rightarrow \frac{1}{6}.(3^2.2).(3).\begin{tikzpicture}[baseline={([yshift=-.5ex]current bounding box.center)}]
    \begin{feynman}
    \vertex (a) at (0,0);
    \node [empty dot, fill=blue] (b) at (0.5,0);
    \node [empty dot, fill=blue] (c) at (1,0);
    \node [empty dot, fill=blue] (d) at (1.35,0.35);
    \node [dot] (e) at (1.35,-0.35);
    \vertex (f) at (1.7,0.7);
    \vertex (g) at (1.7,-0.7);
    \diagram*{(a) -- [red, half right] (b) -- [red] (c) -- [red] (d) -- [red, half right] (f) -- [red, half right] (d) -- [red] (c) -- [red] (e) -- [red, half right] (g) -- [red, half right] (e) -- [red] (c) -- [red] (b) -- [red, half right] (a)};
    \end{feynman}
    \end{tikzpicture},\\
    &\langle(L_2L_3^3)\rangle \rightarrow -\frac{1}{2}(3.2).(3).\begin{tikzpicture}[baseline={([yshift=-.5ex]current bounding box.center)}]
    \begin{feynman}
    \vertex (a) at (0,0);
    \node [empty dot, fill=blue] (b) at (0.5,0);
    \node [empty dot, fill=blue] (c) at (1,0);
    \node [empty dot, fill=blue] (d) at (1.35,0.35);
    \node [dot] (e) at (1.35,-0.35);
    \vertex (f) at (1.7,0.7);
    \vertex (g) at (1.7,-0.7);
    \diagram*{(a) -- [red, half right] (b) -- [red] (c) -- [red] (d) -- [red, half right] (f) -- [red, half right] (d) -- [red] (c) -- [red] (e) -- [red, half right] (g) -- [red, half right] (e) -- [red] (c) -- [red] (b) -- [red, half right] (a)};
    \end{feynman}
    \end{tikzpicture}.\label{3l14c82}
\end{align}
There is only one diagram possible in category (9), that is
\begin{align}
    \begin{tikzpicture}[baseline={([yshift=-.5ex]current bounding box.center)}]
    \begin{feynman}
    \vertex (a) at (0,0);
    \node [empty dot, fill=blue] (b) at (0.5,0);
    \node [dot] (c) at (1,0);
    \node [empty dot, fill=blue] (d) at (1.35,0.35);
    \node [empty dot, fill=blue] (e) at (1.35,-0.35);
    \vertex (f) at (1.7,0.7);
    \vertex (g) at (1.7,-0.7);
    \diagram*{(a) -- [red, half right] (b) -- [red] (c) -- [red] (d) -- [red, half right] (f) -- [red, half right] (d) -- [red] (c) -- [red] (e) -- [red, half right] (g) -- [red, half right] (e) -- [red] (c) -- [red] (b) -- [red, half right] (a)};
    \end{feynman}
    \end{tikzpicture}
\end{align}
which can be ignored based on (\descref{Result 1}) and (\descref{Result 2}).

We can see that all 1PR diagrams cancel away in each category separately.
\subsubsection{$\langle L_3^4\rangle$}
The last term in (\ref{3l1}) contains all the $S_{;ij}$ vertices and hence cannot have any 1PR diagrams according to (\descref{Result 1}) and (\descref{Result 2}) found in Section (\ref{SecV}).

\subsection{$\langle A_3^2A_4\rangle$}\label{SecV2}

There are three vertices in this term out of which three lines emerge from two vertices (the $A_3^2$ term) and four lines emerge from one vertex (the $A_4$ term). It would be a highly tedious job to show one-particle irreducibility term-wise owing to the increasing number of terms beyond $A_3$.  Fortunately, it is not necessary to consider each term in $A_4$ to examine one-particle irreducibility.

1PR diagrams from $\langle A_3^2A_4\rangle$ can be obtained in the following way. We first try to see how four lines coming out from the vertices in $A_4$ join with the three lines coming out from vertices in $A_3$. The product $A_3A_4$ can be represented in a diagrammatic way as follows.
\begin{align}\label{a3a4}
    \underbrace{\Big(\frac{1}{6}\begin{tikzpicture}[baseline={([yshift=-.5ex]current bounding box.center)}]
        \begin{feynman}
            \vertex (a) at (0,0);
            \node [dot](b) at (0.5,0);
            \vertex (c) at (0.85,0.35);
            \vertex (d) at (0.85,-0.35);
            \diagram*{(a) -- [red] (b) -- [red](c)};
            \diagram*{(b) -- [red](d)};
            \end{feynman}
    \end{tikzpicture}-\frac{1}{2}\begin{tikzpicture}[baseline={([yshift=-.5ex]current bounding box.center)}]
        \begin{feynman}
            \vertex (a) at (0,0);
            \node [dot](b) at (0.5,0);
            \vertex (c) at (1,0);
            \diagram*{(a) -- [red, half right] (b) -- [red](c)};
            \diagram*{(b) -- [red, half right](a)};
            \end{feynman}
    \end{tikzpicture} - \begin{tikzpicture}[baseline={([yshift=-.5ex]current bounding box.center)}]
        \begin{feynman}
            \vertex (a) at (0,0);
            \node [empty dot, fill=blue] (b) at (0.5,0);
            \vertex (c) at (0.85,0.35);
            \vertex (d) at (0.85,-0.35);
            \diagram*{(a) -- [red] (b) -- [red](c)};
            \diagram*{(b) -- [red](d)};
            \end{feynman}
    \end{tikzpicture} \ \Big)}_{A_3}\underbrace{\Big(B_1\begin{tikzpicture}[baseline={([yshift=-.5ex]current bounding box.center)}]
        \begin{feynman}
            \node [dot] (a) at (0,0);
            \vertex [dot](b) at (-0.35,0.35);
            \vertex [dot](c) at (-0.35,-0.35);
             \vertex [dot](d) at (0.35,0.35);
              \vertex [dot](e) at (0.35,-0.35);
            \diagram*{(a) -- [red] (b)};
            \diagram*{(a) -- [red] (c)};
            \diagram*{(a) -- [red] (d)};
            \diagram*{(a) -- [red] (e)};
            \end{feynman}
    \end{tikzpicture} + B_2\begin{tikzpicture}[baseline={([yshift=-.5ex]current bounding box.center)}]
        \begin{feynman}
            \vertex (a) at (0,0);
            \node [dot](b) at (0.5,0);
            \vertex (c) at (0.85,0.35);
            \vertex (d) at (0.85,-0.35);
            \diagram*{(a) -- [red, half right] (b) -- [red](c)};
            \diagram*{(d) -- [red] (b) -- [red, half right](a)};
            \end{feynman}
    \end{tikzpicture}\Big)}_{A_4}
\end{align}
where the blue dot in $A_3$ in the expression above represents the $S_{;ij}$ vertex. The term in $A_4$ in (\ref{a3a4}) with coefficient $B_2$ collectively represents the terms $\Gamma^{(1)}_{,i}\sigma^i_{(2)}$ and $S_{,i}C_1^{-1i}{}_{j}[\bar\varphi]\sigma_{(2)}^j$. To understand this we recall that $\Gamma^{(1)}_{,i}$ represents one loop diagram with a vertex. So the product $\Gamma^{(1)}_{,i}\sigma^i_{(2)}$ represents one loop diagram with two lines coming out of a vertex. We also observe that since the connection terms are proportional to Dirac delta functions the Riemann curvature tensor formed from the connection terms is also proportional to the Dirac delta functions and therefore we can write $S_{,i}C_1^{-1i}{}_{j}[\bar\varphi]\sigma_{(2)}^j$ as an integral over a single space-time coordinate following the example in (\ref{example1a}). Looking at the expression for $C_1^{-1i}{}_{j}$ (\ref{Cinv1}) we find that $S_{,i}C_1^{-1i}{}_{j}[\bar\varphi]\sigma_{(2)}^j$ has the same diagrammatic form as $\Gamma^{(1)}_{,i}\sigma^i_{(2)}$. The term in (\ref{a3a4}) with coefficient $B_1$ represents the remaining terms in $A_4$. The product (\ref{a3a4}) can be broken down and simplified as follows.
\begin{align}\label{a3a41}
    &\Big(\frac{1}{6}\begin{tikzpicture}[baseline={([yshift=-.5ex]current bounding box.center)}]
        \begin{feynman}
            \vertex (a) at (0,0);
            \node [dot](b) at (0.5,0);
            \vertex (c) at (0.85,0.35);
            \vertex (d) at (0.85,-0.35);
            \diagram*{(a) -- [red] (b) -- [red](c)};
            \diagram*{(b) -- [red](d)};
            \end{feynman}
    \end{tikzpicture}-\frac{1}{2}\begin{tikzpicture}[baseline={([yshift=-.5ex]current bounding box.center)}]
        \begin{feynman}
            \vertex (a) at (0,0);
            \node [dot](b) at (0.5,0);
            \vertex (c) at (1,0);
            \diagram*{(a) -- [red, half right] (b) -- [red](c)};
            \diagram*{(b) -- [red, half right](a)};
            \end{feynman}
    \end{tikzpicture}\Big)\times B_1 \begin{tikzpicture}[baseline={([yshift=-.5ex]current bounding box.center)}]
        \begin{feynman}
            \node [dot] (a) at (0,0);
            \vertex [dot](b) at (-0.35,0.35);
            \vertex [dot](c) at (-0.35,-0.35);
             \vertex [dot](d) at (0.35,0.35);
              \vertex [dot](e) at (0.35,-0.35);
            \diagram*{(a) -- [red] (b)};
            \diagram*{(a) -- [red] (c)};
            \diagram*{(a) -- [red] (d)};
            \diagram*{(a) -- [red] (e)};
            \end{feynman}
    \end{tikzpicture} = B_1\Big(\frac{c_1}{6}\begin{tikzpicture}[baseline={([yshift=-.5ex]current bounding box.center)}]
        \begin{feynman}
            \vertex (a) at (0,0);
            \node [dot] (b) at (0.5,0);
            \node [dot] (c) at (1,0);
            \diagram*{(a) -- [red] (b) -- [red, half right] (c)};
            \diagram*{(b) -- [red] (c)};
            \diagram*{(c) -- [red, half right] (b)};
            \end{feynman}
    \end{tikzpicture}+\frac{c_2}{6}\begin{tikzpicture}[baseline={([yshift=-.5ex]current bounding box.center)}]
        \begin{feynman}
            \vertex (a) at (0,0);
            \node [dot] (b) at (0.5,0);
            \node [dot] (c) at (1,0);
            \vertex (d) at (1.5,0);
            \diagram*{(a) -- [red] (b) -- [red, half right] (c) -- [red, half right] (d) -- [red, half right] (c) -- [red, half right] (b)};
            \end{feynman}
    \end{tikzpicture} + \frac{3c_3}{6}\begin{tikzpicture}[baseline={([yshift=-.5ex]current bounding box.center)}]
        \begin{feynman}
            \vertex (a) at (0,0);
            \node [dot] (b) at (0.5,0);
            \node [dot] (c) at (1,0);
            \vertex (d) at (1.5,0);
            \vertex (e) at (0.5,0.5);
            \diagram*{(a) -- [red] (b) -- [red] (c) -- [red, half right] (d) -- [red, half right] (c)};
            \diagram*{(b) -- [red, half right] (e) -- [red, half right] (b)};
            \end{feynman}
    \end{tikzpicture} \nonumber\\
    &\hspace{105mm}- \frac{c_3}{2}\begin{tikzpicture}[baseline={([yshift=-.5ex]current bounding box.center)}]
        \begin{feynman}
            \vertex (a) at (0,0);
            \node [dot] (b) at (0.5,0);
            \node [dot] (c) at (1,0);
            \vertex (d) at (1.5,0);
            \vertex (e) at (0.5,0.5);
            \diagram*{(a) -- [red] (b) -- [red] (c) -- [red, half right] (d) -- [red, half right] (c)};
            \diagram*{(b) -- [red, half right] (e) -- [red, half right] (b)};
            \end{feynman}
    \end{tikzpicture}\Big)\\
    &\Big(\frac{1}{6}\begin{tikzpicture}[baseline={([yshift=-.5ex]current bounding box.center)}]
        \begin{feynman}
            \vertex (a) at (0,0);
            \node [dot](b) at (0.5,0);
            \vertex (c) at (0.85,0.35);
            \vertex (d) at (0.85,-0.35);
            \diagram*{(a) -- [red] (b) -- [red](c)};
            \diagram*{(b) -- [red](d)};
            \end{feynman}
    \end{tikzpicture}-\frac{1}{2}\begin{tikzpicture}[baseline={([yshift=-.5ex]current bounding box.center)}]
        \begin{feynman}
            \vertex (a) at (0,0);
            \node [dot](b) at (0.5,0);
            \vertex (c) at (1,0);
            \diagram*{(a) -- [red, half right] (b) -- [red](c)};
            \diagram*{(b) -- [red, half right](a)};
            \end{feynman}
    \end{tikzpicture}\Big)\times B_2\begin{tikzpicture}[baseline={([yshift=-.5ex]current bounding box.center)}]
        \begin{feynman}
            \vertex (a) at (0,0);
            \node [dot](b) at (0.5,0);
            \vertex (c) at (0.85,0.35);
            \vertex (d) at (0.85,-0.35);
            \diagram*{(a) -- [red, half right] (b) -- [red](c)};
            \diagram*{(d) -- [red] (b) -- [red, half right](a)};
            \end{feynman}
    \end{tikzpicture} = B_2\Big(\frac{c_4}{6}\begin{tikzpicture}[baseline={([yshift=-.5ex]current bounding box.center)}]
        \begin{feynman}
            \vertex (a) at (0,0);
            \node [dot] (b) at (0.5,0);
            \node [dot] (c) at (1,0);
            \vertex (d) at (1.5,0);
            \diagram*{(a) -- [red] (b) -- [red, half right] (c) -- [red, half right] (d) -- [red, half right] (c) -- [red, half right] (b)};
            \end{feynman}
    \end{tikzpicture} + \frac{3c_5}{6}\begin{tikzpicture}[baseline={([yshift=-.5ex]current bounding box.center)}]
        \begin{feynman}
            \vertex (a) at (0,0);
            \node [dot] (b) at (0.5,0);
            \node [dot] (c) at (1,0);
            \vertex (d) at (1.5,0);
            \vertex (e) at (0.5,0.5);
            \diagram*{(a) -- [red] (b) -- [red] (c) -- [red, half right] (d) -- [red, half right] (c)};
            \diagram*{(b) -- [red, half right] (e) -- [red, half right] (b)};
            \end{feynman}
    \end{tikzpicture} - \frac{c_5}{2}\begin{tikzpicture}[baseline={([yshift=-.5ex]current bounding box.center)}]
        \begin{feynman}
            \vertex (a) at (0,0);
            \node [dot] (b) at (0.5,0);
            \node [dot] (c) at (1,0);
            \vertex (d) at (1.5,0);
            \vertex (e) at (0.5,0.5);
            \diagram*{(a) -- [red] (b) -- [red] (c) -- [red, half right] (d) -- [red, half right] (c)};
            \diagram*{(b) -- [red, half right] (e) -- [red, half right] (b)};
            \end{feynman}
    \end{tikzpicture}\label{a3a42}\\
    &-\begin{tikzpicture}[baseline={([yshift=-.5ex]current bounding box.center)}]
        \begin{feynman}
            \vertex (a) at (0,0);
            \node [empty dot, fill=blue] (b) at (0.5,0);
            \vertex (c) at (0.85,0.35);
            \vertex (d) at (0.85,-0.35);
            \diagram*{(a) -- [red] (b) -- [red](c)};
            \diagram*{(b) -- [red](d)};
            \end{feynman}
    \end{tikzpicture}\times B_1 \begin{tikzpicture}[baseline={([yshift=-.5ex]current bounding box.center)}]
        \begin{feynman}
            \node [dot] (a) at (0,0);
            \vertex [dot](b) at (-0.35,0.35);
            \vertex [dot](c) at (-0.35,-0.35);
             \vertex [dot](d) at (0.35,0.35);
              \vertex [dot](e) at (0.35,-0.35);
            \diagram*{(a) -- [red] (b)};
            \diagram*{(a) -- [red] (c)};
            \diagram*{(a) -- [red] (d)};
            \diagram*{(a) -- [red] (e)};
            \end{feynman}
    \end{tikzpicture} = B_1\Big(- c_1\begin{tikzpicture}[baseline={([yshift=-.5ex]current bounding box.center)}]
        \begin{feynman}
            \vertex (a) at (0,0);
            \node [dot] (b) at (0.5,0);
            \node [empty dot, fill=blue] (c) at (1,0);
            \diagram*{(a) -- [red] (b) -- [red, half right] (c)};
            \diagram*{(b) -- [red] (c)};
            \diagram*{(c) -- [red, half right] (b)};
            \end{feynman}
    \end{tikzpicture}-c_2\begin{tikzpicture}[baseline={([yshift=-.5ex]current bounding box.center)}]
        \begin{feynman}
            \vertex (a) at (0,0);
            \node [empty dot, fill=blue] (b) at (0.5,0);
            \node [dot] (c) at (1,0);
            \vertex (d) at (1.5,0);
            \diagram*{(a) -- [red] (b) -- [red, half right] (c) -- [red, half right] (d) -- [red, half right] (c) -- [red, half right] (b)};
            \end{feynman}
    \end{tikzpicture}  - c_3\begin{tikzpicture}[baseline={([yshift=-.5ex]current bounding box.center)}]
        \begin{feynman}
            \vertex (a) at (0,0);
            \node [dot] (b) at (0.5,0);
            \node [empty dot, fill=blue] (c) at (1,0);
            \vertex (d) at (1.5,0);
            \vertex (e) at (0.5,0.5);
            \diagram*{(a) -- [red] (b) -- [red] (c) -- [red, half right] (d) -- [red, half right] (c)};
            \diagram*{(b) -- [red, half right] (e) -- [red, half right] (b)};
            \end{feynman}
    \end{tikzpicture}\Big)\label{a3a43}\\
    &-\begin{tikzpicture}[baseline={([yshift=-.5ex]current bounding box.center)}]
        \begin{feynman}
            \vertex (a) at (0,0);
            \node [empty dot, fill=blue] (b) at (0.5,0);
            \vertex (c) at (0.85,0.35);
            \vertex (d) at (0.85,-0.35);
            \diagram*{(a) -- [red] (b) -- [red](c)};
            \diagram*{(b) -- [red](d)};
            \end{feynman}
    \end{tikzpicture}\times B_2\begin{tikzpicture}[baseline={([yshift=-.5ex]current bounding box.center)}]
        \begin{feynman}
            \vertex (a) at (0,0);
            \node [dot](b) at (0.5,0);
            \vertex (c) at (0.85,0.35);
            \vertex (d) at (0.85,-0.35);
            \diagram*{(a) -- [red, half right] (b) -- [red](c)};
            \diagram*{(d) -- [red] (b) -- [red, half right](a)};
            \end{feynman}
    \end{tikzpicture} = B_2\Big(-c_4\begin{tikzpicture}[baseline={([yshift=-.5ex]current bounding box.center)}]
        \begin{feynman}
            \vertex (a) at (0,0);
            \node [empty dot, fill=blue] (b) at (0.5,0);
            \node [dot] (c) at (1,0);
            \vertex (d) at (1.5,0);
            \diagram*{(a) -- [red] (b) -- [red, half right] (c) -- [red, half right] (d) -- [red, half right] (c) -- [red, half right] (b)};
            \end{feynman}
    \end{tikzpicture}  - c_5\begin{tikzpicture}[baseline={([yshift=-.5ex]current bounding box.center)}]
        \begin{feynman}
            \vertex (a) at (0,0);
            \node [dot] (b) at (0.5,0);
            \node [empty dot, fill=blue] (c) at (1,0);
            \vertex (d) at (1.5,0);
            \vertex (e) at (0.5,0.5);
            \diagram*{(a) -- [red] (b) -- [red] (c) -- [red, half right] (d) -- [red, half right] (c)};
            \diagram*{(b) -- [red, half right] (e) -- [red, half right] (b)};
            \end{feynman}
    \end{tikzpicture}\Big)\label{a3a44}
\end{align}
where the coefficients $c_1$, $c_2$, $c_3$, $c_4$, and $c_5$ are real numbers that may be considered to be the number of ways to form the diagrams. We have explicitly shown that irrespective of the value of constants $c_3$ and $c_5$, the number of ways to form 1PR diagrams in $L_1A_4$, $\Big(\frac{1}{6}\begin{tikzpicture}[baseline={([yshift=-.5ex]current bounding box.center)}]
        \begin{feynman}
            \vertex (a) at (0,0);
            \node [dot](b) at (0.5,0);
            \vertex (c) at (0.85,0.35);
            \vertex (d) at (0.85,-0.35);
            \diagram*{(a) -- [red] (b) -- [red](c)};
            \diagram*{(b) -- [red](d)};
            \end{feynman}
    \end{tikzpicture} \ \Big)\Big(B_1\begin{tikzpicture}[baseline={([yshift=-.5ex]current bounding box.center)}]
        \begin{feynman}
            \node [dot] (a) at (0,0);
            \vertex [dot](b) at (-0.35,0.35);
            \vertex [dot](c) at (-0.35,-0.35);
             \vertex [dot](d) at (0.35,0.35);
              \vertex [dot](e) at (0.35,-0.35);
            \diagram*{(a) -- [red] (b)};
            \diagram*{(a) -- [red] (c)};
            \diagram*{(a) -- [red] (d)};
            \diagram*{(a) -- [red] (e)};
            \end{feynman}
    \end{tikzpicture} + B_2\begin{tikzpicture}[baseline={([yshift=-.5ex]current bounding box.center)}]
        \begin{feynman}
            \vertex (a) at (0,0);
            \node [dot](b) at (0.5,0);
            \vertex (c) at (0.85,0.35);
            \vertex (d) at (0.85,-0.35);
            \diagram*{(a) -- [red, half right] (b) -- [red](c)};
            \diagram*{(d) -- [red] (b) -- [red, half right](a)};
            \end{feynman}
    \end{tikzpicture}\Big)$, is always thrice the number of ways to form 1PR diagrams in $L_2A_4$, $\Big(-\frac{1}{2}\begin{tikzpicture}[baseline={([yshift=-.5ex]current bounding box.center)}]
        \begin{feynman}
            \vertex (a) at (0,0);
            \node [dot](b) at (0.5,0);
            \vertex (c) at (1,0);
            \diagram*{(a) -- [red, half right] (b) -- [red](c)};
            \diagram*{(b) -- [red, half right](a)};
            \end{feynman}
    \end{tikzpicture} \ \Big)\Big(B_1\begin{tikzpicture}[baseline={([yshift=-.5ex]current bounding box.center)}]
        \begin{feynman}
            \node [dot] (a) at (0,0);
            \vertex [dot](b) at (-0.35,0.35);
            \vertex [dot](c) at (-0.35,-0.35);
             \vertex [dot](d) at (0.35,0.35);
              \vertex [dot](e) at (0.35,-0.35);
            \diagram*{(a) -- [red] (b)};
            \diagram*{(a) -- [red] (c)};
            \diagram*{(a) -- [red] (d)};
            \diagram*{(a) -- [red] (e)};
            \end{feynman}
    \end{tikzpicture} + B_2\begin{tikzpicture}[baseline={([yshift=-.5ex]current bounding box.center)}]
        \begin{feynman}
            \vertex (a) at (0,0);
            \node [dot](b) at (0.5,0);
            \vertex (c) at (0.85,0.35);
            \vertex (d) at (0.85,-0.35);
            \diagram*{(a) -- [red, half right] (b) -- [red](c)};
            \diagram*{(d) -- [red] (b) -- [red, half right](a)};
            \end{feynman}
    \end{tikzpicture}\Big)$. This is because a loop can be formed in three ways by joining the lines in $L_1$. As a result, all the 1PR diagrams cancel away in (\ref{a3a41}) and (\ref{a3a42}). In the last two expressions because the blue vertex carries $S_{;ij}$, using (\descref{Result 1}) and (\descref{Result 2}) we find that the 1PR diagrams either regularise to zero or turn into 1PI diagrams. The result of this discussion is that the product (\ref{a3a4}) does not contain any 1PR diagrams.

Based on what we have found, we can say that the product (\ref{a3a4}) can be written as
\begin{align}\label{a3a4compact}
    \begin{tikzpicture}[square/.style={regular polygon,regular polygon sides=4}]
    \begin{feynman}
         \vertex at (0,0) (a);
        \node at (1.5,0) [square,inner sep=-0.3em,draw] (b) {$A_3A_4$};
        \diagram*{ (a) -- [red] (b)};
        \draw[fill=black] (1.0,0) circle(0.75mm);
    \end{feynman}
    \end{tikzpicture},
\end{align}
where the box represents all possible 1PI diagrams that one can form by joining the lines coming out from the vertices in $A_3$ and $A_4$ with the additional 1PR diagrams (with blue vertices) that turn into 1PI diagrams. On multiplying (\ref{a3a4compact}) further with $A_3$ we find
\begin{align}\label{a3a4a3}
   \Big(\frac{1}{6}\begin{tikzpicture}[baseline={([yshift=-.5ex]current bounding box.center)}]
        \begin{feynman}
            \vertex (a) at (0,0);
            \node [dot](b) at (0.5,0);
            \vertex (c) at (0.85,0.35);
            \vertex (d) at (0.85,-0.35);
            \diagram*{(a) -- [red] (b) -- [red](c)};
            \diagram*{(b) -- [red](d)};
            \end{feynman}
    \end{tikzpicture}-\frac{1}{2}\begin{tikzpicture}[baseline={([yshift=-.5ex]current bounding box.center)}]
        \begin{feynman}
            \vertex (a) at (0,0);
            \node [dot](b) at (0.5,0);
            \vertex (c) at (1,0);
            \diagram*{(a) -- [red, half right] (b) -- [red](c)};
            \diagram*{(b) -- [red, half right](a)};
            \end{feynman}
    \end{tikzpicture} - \begin{tikzpicture}[baseline={([yshift=-.5ex]current bounding box.center)}]
        \begin{feynman}
            \vertex (a) at (0,0);
            \node [empty dot, fill=blue] (b) at (0.5,0);
            \vertex (c) at (0.85,0.35);
            \vertex (d) at (0.85,-0.35);
            \diagram*{(a) -- [red] (b) -- [red](c)};
            \diagram*{(b) -- [red](d)};
            \end{feynman}
    \end{tikzpicture} \ \Big)\times \begin{tikzpicture}[baseline={([yshift=-.5ex]current bounding box.center)}, square/.style={regular polygon,regular polygon sides=4}]
    \begin{feynman}
         \vertex at (0,0) (a);
        \node at (1.3,0) [square,inner sep=-0.3em,draw] (b) {$A_3A_4$};
        \diagram*{ (a) -- [red] (b)};
        \draw[fill=black] (0.8,0) circle(0.75mm);
    \end{feynman}
    \end{tikzpicture} = 0 -\begin{tikzpicture}[baseline={([yshift=-.5ex]current bounding box.center)}, square/.style={regular polygon,regular polygon sides=4}]
    \begin{feynman}
    \vertex at (-0.5,0) (c);
         \node [empty dot, fill = blue]at (0,0) (a);
        \node at (1.3,0) [square,inner sep=-0.3em,draw] (b) {$A_3A_4$};
        \diagram*{ (c) -- [red, half right] (a) -- [red] (b)};
        \draw[fill=black] (0.8,0) circle(0.75mm);
        \diagram*{(a) -- [red, half right] (c)};
    \end{feynman}
    \end{tikzpicture}.
\end{align}

Again it is clear that since there are three ways to form a loop by joining the lines in $L_1$ all 1PR diagrams cancel away when $L_1$ and $L_2$ are multiplied by (\ref{a3a4compact}) in (\ref{a3a4a3}). The remaining diagram on the RHS of (\ref{a3a4a3}) either regularises to zero or turns into a 1PI diagram using (\descref{Result 1}) and (\descref{Result 2}). 

Thus, we have shown that $\langle A_3^2A_4\rangle$ only contains 1PI diagrams.

\subsection{$\langle A_3A_5\rangle$}\label{SecV3}
This term contains two vertices; one with three lines and one with five. The only way 1PI diagrams can form from this term is
\begin{align}\label{a3a5}
   \Big(\frac{1}{6}\begin{tikzpicture}[baseline={([yshift=-.5ex]current bounding box.center)}]
        \begin{feynman}
            \vertex (a) at (0,0);
            \node [dot](b) at (0.5,0);
            \vertex (c) at (0.85,0.35);
            \vertex (d) at (0.85,-0.35);
            \diagram*{(a) -- [red] (b) -- [red](c)};
            \diagram*{(b) -- [red](d)};
            \end{feynman}
    \end{tikzpicture}-\frac{1}{2}\begin{tikzpicture}[baseline={([yshift=-.5ex]current bounding box.center)}]
        \begin{feynman}
            \vertex (a) at (0,0);
            \node [dot](b) at (0.5,0);
            \vertex (c) at (1,0);
            \diagram*{(a) -- [red, half right] (b) -- [red](c)};
            \diagram*{(b) -- [red, half right](a)};
            \end{feynman}
    \end{tikzpicture} - \begin{tikzpicture}[baseline={([yshift=-.5ex]current bounding box.center)}]
        \begin{feynman}
            \vertex (a) at (0,0);
            \node [empty dot, fill=blue] (b) at (0.5,0);
            \vertex (c) at (0.85,0.35);
            \vertex (d) at (0.85,-0.35);
            \diagram*{(a) -- [red] (b) -- [red](c)};
            \diagram*{(b) -- [red](d)};
            \end{feynman}
    \end{tikzpicture} \ \Big)\times \begin{tikzpicture}[baseline={([yshift=-.5ex]current bounding box.center)}, square/.style={regular polygon,regular polygon sides=4}]
    \begin{feynman}
         \vertex at (0,0) (a);
        \node at (1.3,0) [square,inner sep=0em,draw] (b) {$A_5$};
        \diagram*{ (a) -- [red] (b)};
        \draw[fill=black] (0.98,0) circle(0.75mm);
    \end{feynman}
    \end{tikzpicture} = 0 -\begin{tikzpicture}[baseline={([yshift=-.5ex]current bounding box.center)}, square/.style={regular polygon,regular polygon sides=4}]
    \begin{feynman}
    \vertex at (-0.5,0) (c);
         \node [empty dot, fill = blue]at (0,0) (a);
        \node at (1.3,0) [square,inner sep=0em,draw] (b) {$A_5$};
        \diagram*{ (c) -- [red, half right] (a) -- [red] (b)};
        \draw[fill=black] (0.98,0) circle(0.75mm);
        \diagram*{(a) -- [red, half right] (c)};
    \end{feynman}
    \end{tikzpicture}, 
\end{align}

where the box represents 1PI diagrams formed by joining four lines out of five coming out of vertices in $A_5$. The discussion on one-particle irreducibility for this term is identical to that for $\langle A_3^2A_4\rangle$ as done in the preceding sub-section below (\ref{a3a4compact}), from which it follows that $\langle A_3A_5\rangle$ does not contain any 1PR diagrams.

\subsection{$\langle A_4^2 \rangle$}\label{SecV4}
This term contains two vertices with four lines coming out from each of them. It is easy to see that this term cannot form any 1PR diagram.

The results from (\ref{SecV1}), (\ref{SecV2}), (\ref{SecV3}), and (\ref{SecV4}) show that VDEA up to three loops can be written as a sum of one-particle irreducible diagrams.

%% file: Conclusion.tex
In this paper, we discussed VDEA and its one-particle irreducibility for non-gauge theories. We described two methods to compute VDEA, (1) where the limit $\varphi_*^i=\bar\varphi^i$ is taken after computing the effective action and (2) where the limit is taken before. Method (1) involves solving the equation (\ref{vdea1}) while using $\sigma^i[\varphi_*,\varphi]$ as the variable for the path integral whereas method (2) requires us to use the original fields $\varphi^i$ as the variable for the path integral. It is easy to show one-particle irreducibility when method (1) is employed since we must solve only one equation (\ref{vdea1}). However, one-particle irreducibility of VDEA is not easily seen if method (2) is employed because it requires coupled differential equations (\ref{vdea2re}) and (\ref{Ceq}) to be solved. As pointed out in \cite{Rebhan:1986wp, Rebhan:1987cd}, we saw that the extra term $C^{-1i}{}_j$ is crucial for symmetrization of indices in the vertex $S_{;ijk}$ to happen leading to cancellation of 1PR diagrams. However, as shown in Sections (\ref{SecIV}) and (\ref{SecV}) the normal coordinate expansion of $\sigma^i[\bar\varphi,\varphi]$ also throws in extra terms which need to be checked for one-particle irreducibility. Fortunately, the extra terms involved new vertices proportional to $S_{;ij}$ which is the inverse of Green's functions. This fact was crucial to derive two results regarding 1PR diagrams that involved vertices with $S_{;ij}$; (1) certain 1PR diagrams could be regularised to zero when dimensional regularisation is used (See \descref{Result 1}), and (2) certain 1PR diagrams turn into 1PI diagrams (\descref{Result 2}). These results were then used at three-loop order to establish one-particle irreducibility of VDEA for non-gauge theories. 

Beyond three-loop order, we expect symmetrization of the indices in the vertex $S_{;ijkl}$ to happen through the term $C_2^{-1i}{}_j$ in $A_4$. The non-trivial term in VDEA to deal with at four-loop order would be $\langle A_4^2\rangle$ where it would be necessary to know the exact form of $\Gamma^{(2)}_{,i}
$ and $C_2^{-1i}{}_j$. This is unlike the case for two-loop and three-loop VDEA, where the exact form of the terms in $A_4$ did not matter (See Sections (\ref{SecV2}) and (\ref{SecV4})). Furthermore, considering the term $\langle A_4^2\rangle$ the expansion of $\sigma^i[\bar\varphi,\varphi]$ in powers of $\eta^i$ (\ref{sigmaexp}) brings in new terms important at four-loop order which does not carry the factor $S_{;ij}$ such as $(S_{;(ijk)}\eta^i\eta^j\sigma^k_{(2)})^2$). The results (\descref{Result 1}) and (\descref{Result 2}) would not hold there as they did for the terms proportional to $S_{;ij}$ at two-loop and three-loop order. As such, explicit calculations must be performed to ascertain that VDEA is one-particle irreducible beyond three-loop.

%% file: Appendix.tex
\section{DeWitt's Notation}\label{notation}
DeWitt's notation is a shorthand notation which we will be following in this paper. A brief description of DeWitt's notation is as follows.

The symbols used to represent the properties of a field are labeled as $\varphi^i$, where the discrete field index and the field's spacetime argument are condensed into the single label $i$. For instance, if the field is a scalar field, then $\varphi^i$ is equivalent to $\phi(x)$, and if it is a vector field, then $\varphi^i$ is equivalent to $A_\mu(x)$, while for a second-rank tensor field, $\varphi^i$ is equivalent to $g_{\mu\nu}(x)$. Additionally, the following summation convention is used in $n$ dimensions:

\begin{equation}\label{2}
    \varphi^i B_{ij} \varphi^j = \int d^nx\int d^nx' \varphi^I(x) B_{IJ}(x,x') \varphi^J(x'),
\end{equation}

where capital Latin letters $(I, J,...)$ are used as a placeholder for conventional field indices. Thus, the small Latin indices such as `$i$' stands for the pair $(I,x)$.

In curved spacetime the following relation holds,

\begin{align}\label{3}
     \varphi^I(x) &=\int dv_x'\delta(x,x')\varphi^I(x')\\
     &= \int d^nx' |g(x')|^{1/2}\delta(x,x') \varphi^I(x')\\
     & = \int d^nx' \tilde\delta(x,x') \varphi^I(x').
\end{align}
where $dv_x$ is the invariant space-time volume element defined by $dv_x = d^nx|g(x)|^{1/2}$ and $|g(x)|$ is the determinant of the metric $g_{\mu\nu}(x)$. In the third line, we have defined,
 
\begin{equation}\label{4}
    \tilde\delta(x,x') = |g(x')|^{1/2}\delta(x,x'),
\end{equation}

where $\delta(x,x')$ is the conventional bi-scalar Dirac $\delta$-distribution.

\section{Calculation of $C^{-1i}{}_j$}\label{calcCinv}
The method (2) described in Sec. (\ref{SecIII}) requires solving two coupled equations to obtain VDEA. These are;
\begin{align}
    &e^{\frac{i}{\hbar}\Gamma[\bar\varphi]} = \int \mathcal{D}\varphi \ \exp{\frac{i}{\hbar}\left\{S[\varphi]-\sigma^i[\bar\varphi,\varphi]C^{-1j}{}_{,i}[\bar\varphi]\Gamma_{,j}[\bar\varphi]\right\}}\label{vdea2reapp}\\
    &C^i{}_j[\bar\varphi]=e^{-\frac{i}{\hbar}\Gamma[\bar\varphi]}\int \mathcal{D}\varphi \ \sigma^i{}_{;j}[\bar\varphi,\varphi]\exp{\frac{i}{\hbar}\left\{S[\varphi]-\sigma^i[\bar\varphi,\varphi]C^{-1j}{}_{,i}[\bar\varphi]\Gamma_{,j}[\bar\varphi]\right\}}\label{Ceqapp}
\end{align}
To solve these equations we require the expression for $\sigma^i{}_{;j}[\bar\varphi,\varphi]$ which can be found using (\ref{sigmaexp});
\begin{align}\label{sigmacov}
    \sigma^i{}_{;j} = \delta^i_j - \frac{1}{3}R^i{}_{kjl}[\bar\varphi]\eta^k\eta^l + \mathcal{O}(\eta^3)
\end{align}
where $R^i{}_{kjl}[\bar\varphi]$ is the Riemann tensor given by,

\begin{align}
    R^i{}_{kjl}[\bar\varphi] = \Gamma^i_{lk,j}[\bar\varphi]-\Gamma^i_{jk,l}[\bar\varphi] + \Gamma^i_{jm}[\bar\varphi]\Gamma^m_{lk}[\bar\varphi] - \Gamma^i_{lm}[\bar\varphi]\Gamma^m_{jk}[\bar\varphi].
\end{align}

The expression (\ref{sigmacov}) holds also for the quantity $v^i = \sigma^i[\varphi_*,\bar\varphi]$ in powers of $(\bar\varphi^i-\varphi^i_*)$. It is easy to see that in the limit $\varphi^i_* = \bar\varphi^i$ we obtain,

\begin{align}\label{vij}
    \lim_{\varphi^i_* \rightarrow \bar\varphi^i}v^i_{;j}[\varphi_*,\bar\varphi] = \delta^i_j
\end{align}

To compute VDEA up to three loops we require the expression for $\sigma^i_{;j}$ up to fourth order in $\eta^i$ (or second order in $\hbar$). But to examine the one-particle irreducibility of VDEA up to three loops (\ref{sigmacov}) is sufficient as is evident from Sections (\ref{SecV3}) and (\ref{SecV4}). Thus we require only the first two terms in (\ref{Cinvexp}) for our purpose.

Let us find $C^i{}_j$ up to order $\hbar$. $C^i{}_j$ may be written as an infinite series in powers of $\hbar$;

\begin{align}
    C^i{}_j[\bar\varphi] = \sum_{n=0}^\infty\hbar^nC_n^i{}_j[\bar\varphi]
\end{align}

If we substitute $\sigma^i_{;j}[\bar\varphi,\varphi]$ with $\delta^i_j$ then we find;
\begin{align}
    C_0^i{}_j = \delta^i_j
\end{align}

 At order $\hbar$ we substitute $\sigma^i{}_{;j} = -\frac{1}{3}R^i{}_{kjl}\eta^k\eta^l$ in (\ref{Ceqapp}) to obtain;

 \begin{align}
C_1^i{}_j[\bar\varphi] = \frac{\int \mathcal{D}\varphi \Big(-\frac{\hbar}{3}R^i{}_{kjl}[\bar\varphi]\eta^k\eta^l\Big)\exp{\frac{i}{\hbar}\Big\{\medmath{\splitfrac{S[\bar\varphi]+ \sum_{n=0}^\infty \frac{\delta^nS[\bar\varphi]}{\delta\sigma^{i_1}\delta\sigma^{i_2}...\delta\sigma^{i_n}}\sigma^{i_1}\sigma^{i_2}...\sigma^{i_n}}{ \hspace{20mm}-\sigma^i\sum_{k=0}^\infty\hbar^kC_k^{-1i}{}_{j}[\bar\varphi]\sum_{m=0}^\infty\hbar^m\Gamma^{(m)}[\bar\varphi]\Big\}}\rule[-1.5ex]{0pt}{1ex}}}}{\int \mathcal{D}\varphi \ \exp{\frac{i}{\hbar}\Big\{\medmath{\splitfrac{S[\bar\varphi]+ \sum_{n=0}^\infty \frac{\delta^nS[\bar\varphi]}{\delta\sigma^{i_1}\delta\sigma^{i_2}...\delta\sigma^{i_n}}\sigma^{i_1}\sigma^{i_2}...\sigma^{i_n}}{ \hspace{20mm}-\sigma^i\sum_{k=0}^\infty\hbar^kC_k^{-1i}{}_{j}[\bar\varphi]\sum_{m=0}^\infty\hbar^m\Gamma^{(m)}[\bar\varphi]\Big\}}\rule[-1.5ex]{0pt}{1ex}}}}
 \end{align}

 The exponentials must be expanded up to zeroth order in $\hbar$. The $S[\bar\varphi]$ cancels away from the numerator and the denominator and all interaction terms must be dropped to find,
 \begin{align}
     C_1^i{}_j[\bar\varphi] = \frac{\int \mathcal{D}\varphi \Big(-\frac{\hbar}{3}R^i{}_{kjl}[\bar\varphi]\eta^k\eta^l\Big)\exp{\{\frac{i}{2}S_{;ij}\eta^i\eta^j}\}}{\mathcal{D}\varphi \ \exp{\{\frac{i}{2}S_{;ij}\eta^i\eta^j}\}} = -\frac{\hbar}{3}R^i{}_{kjl}[\bar\varphi]\langle\eta^k\eta^l\rangle
 \end{align}
 Inverting $C^i{}_j[\bar\varphi]$ we obtain up to first order in $\hbar$,
 \begin{align}\label{Cinv0}
     &C_0^{-1i}{}_j[\bar\varphi] = \delta^i_j \\
     &C_1^{-1i}{}_j[\bar\varphi] = \frac{\hbar}{3}R^i{}_{kjl}\langle\eta^k\eta^l\rangle\label{Cinv1}
 \end{align}

%% file: main.bbl
\providecommand{\href}[2]{#2}\begingroup\raggedright\begin{thebibliography}{10}

\bibitem{tomsbook}
L.~Parker and D.~Toms, \emph{\href{https://www.cambridge.org/in/academic/subjects/physics/theoretical-physics-and-mathematical-physics/quantum-field-theory-curved-spacetime-quantized-fields-and-gravity?format=HB&isbn=9780521877879}{Quantum field theory in curved spacetime: quantized fields and gravity}}, Cambridge university press (2009).

\bibitem{Odintsovbook}
I.L.~Buchbinder, S.D.~Odintsov and I.L.~Shapiro, \emph{\href{https://www.routledge.com/Effective-Action-in-Quantum-Gravity/Buchbinder-Odintsov-Shapiro/p/book/9780750301220}{Effective action in quantum gravity}}, Routledge (2017).

\bibitem{Vilkovisky:1984st}
G.A.~Vilkovisky, \emph{{The Unique Effective Action in Quantum Field Theory}}, \href{https://doi.org/10.1016/0550-3213(84)90228-1}{\emph{Nucl. Phys. B} {\bfseries 234} (1984) 125}.

\bibitem{Rebhan:1986wp}
A.~Rebhan, \emph{{The Vilkovisky-de Witt Effective Action and Its Application to {Yang-Mills} Theories}}, \href{https://doi.org/10.1016/0550-3213(87)90241-0}{\emph{Nucl. Phys. B} {\bfseries 288} (1987) 832}.

\bibitem{Rebhan:1987cd}
A.~Rebhan, \emph{{Feynman Rules and S Matrix Equivalence of the Vilkovisky-de Witt Effective Action}}, \href{https://doi.org/10.1016/0550-3213(88)90005-3}{\emph{Nucl. Phys. B} {\bfseries 298} (1988) 726}.

\bibitem{Burgess:1987zi}
C.P.~Burgess and G.~Kunstatter, \emph{{On the Physical Interpretation of the Vilkovisky-de Witt Effective Action}}, \href{https://doi.org/10.1142/S0217732387001117}{\emph{Mod. Phys. Lett. A} {\bfseries 2} (1987) 875}.

\bibitem{Batalin:1987fw}
B.S.~DeWitt, \emph{{The effective action, in Quantum Field Theory and Quantum Statistics, essays in honor of the sixtieth birthday of E.S. Fradkin, Vol. 1: Quantum Statistics and methods of Field Theory}}, C.J. Isham, I.A. Batalin and G.A. Vilkovisky eds., Hilger, Bristol, U.K (1987).

\bibitem{paper1}
S.~Aashish, S.~Panda, A.A.~Tinwala and A.~Vidyarthi, \emph{{Covariant effective action for scalar-tensor theories of gravity}}, \href{https://doi.org/10.1088/1475-7516/2021/10/006}{\emph{JCAP} {\bfseries 10} (2021) 006} [\href{https://arxiv.org/abs/2104.12713}{{\ttfamily 2104.12713}}].

\bibitem{paper2}
S.~Panda, A.A.~Tinwala and A.~Vidyarthi, \emph{{Covariant effective action for generalized Proca theories}}, \href{https://doi.org/10.1088/1475-7516/2022/01/062}{\emph{JCAP} {\bfseries 01} (2022) 062} [\href{https://arxiv.org/abs/2112.04391}{{\ttfamily 2112.04391}}].

\bibitem{paper3}
S.~Panda, A.A.~Tinwala and A.~Vidyarthi, \emph{{Local momentum space: scalar field and gravity}}, \href{https://doi.org/10.1088/1361-6382/ad04b2}{\emph{Class. Quant. Grav.} {\bfseries 40} (2023) 235001} [\href{https://arxiv.org/abs/2205.12842}{{\ttfamily 2205.12842}}].

\bibitem{cho1989vilkovisky}
H.T.~Cho, \emph{\href{https://journals.aps.org/prd/abstract/10.1103/PhysRevD.40.3302}{Vilkovisky-DeWitt effective potential for Einstein gravity coupled to scalars}}, {\emph{Physical Review D} {\bfseries 40} (1989) 3302}.

\bibitem{cho1991vilkovisky}
H.T.~Cho, \emph{\href{https://journals.aps.org/prd/abstract/10.1103/PhysRevD.43.1859}{Vilkovisky-DeWitt effective potential for higher-derivative gravity coupled to scalars}}, {\emph{Physical Review D} {\bfseries 43} (1991) 1859}.

\bibitem{odintsov1993gaugeh}
S.~Odintsov and I.~Shevchenko, \emph{\href{https://onlinelibrary.wiley.com/doi/abs/10.1002/prop.2190410803}{Gauge-Invariant and Gauge-Fixing Independent Effective Action in One-Loop Quantum Gravity}}, {\emph{Fortschritte der Physik/Progress of Physics} {\bfseries 41} (1993) 719}.

\bibitem{xTensor}
J.M.~Martín-García, ``\href{http://xact.es/xTensor/}{xTensor: Fast abstract tensor computer algebra}.''

\bibitem{xPert}
J.M.M.-G.~David~Brizuela and G.A.M.~Marugán, ``\href{http://xact.es/xPert/index.html}{xPert: Computer algebra for metric perturbation theory}.''

\end{thebibliography}\endgroup
